\documentclass[acmsmall]{acmart}
\AtBeginDocument{%
  }

\usepackage{algorithm}
\usepackage{algpseudocode}
\usepackage{tabularx}
\usepackage{array}
\usepackage{booktabs}
\usepackage{makecell}
\usepackage{multirow} 
\usepackage{threeparttable}

\setcopyright{acmlicensed}
\copyrightyear{2018}
\acmYear{2018}
\acmDOI{XXXXXXX.XXXXXXX}

\acmJournal{JACM}
\acmVolume{37}
\acmNumber{4}
\acmArticle{111}
\acmMonth{8}

\begin{document}

\title{MemConflict: Evaluating Long-Term Memory Systems Under Memory Conflicts}

\author{Zhen Tao}
\affiliation{
  \institution{Renmin University of China}
  \city{Beijing}
  \country{China}
}
\email{taozhen@ruc.edu.cn}

\author{Jinxiang Zhao}
\affiliation{
  \institution{Renmin University of China}
  \city{Beijing}
  \country{China}
}
\email{2025104017@ruc.edu.cn}

\author{Peng Liu}
\affiliation{
  \institution{Renmin University of China}
  \city{Beijing}
  \country{China}
}
\email{cs_liupeng@ruc.edu.cn}

\author{Dinghao Xi}
\authornote{Corresponding authors.}
\affiliation{%
  \institution{Shanghai University of Finance and Economics}
  \city{Shanghai}
  \country{China}}
\email{xidinghao@mail.sufe.edu.cn}

\author{Yanfang Chen}
\authornotemark[1]
\affiliation{%
  \institution{Renmin University of China}
  \city{Beijing}
  \country{China}
}
\email{cyf@ruc.edu.cn}

\author{Wei Xu}
\affiliation{
  \institution{Renmin University of China}
  \city{Beijing}
  \country{China}}
\email{weixu@ruc.edu.cn}

\author{Zhiyu Li}
\affiliation{%
  \institution{MemTensor (Shanghai) Technology}
  \city{Shanghai}
  \country{China}
}
\email{lizy@iaar.ac.cn}

\renewcommand{\shortauthors}{Tao et al.}

\begin{abstract}
Long-term memory systems enable conversational agents based on large language models (LLMs) to retain, retrieve, and apply user-specific information across multi-session interactions. However, existing evaluations mainly assess outcome-level performance or temporal updating, providing limited insight into how systems retrieve and rank temporally valid, factually correct, and contextually applicable memory evidence under conflicting alternatives. To address this gap, we propose \textbf{MemConflict}, a diagnostic framework that treats memory validity as a query-conditioned fitness-for-use problem. MemConflict formalizes \textbf{dynamic}, \textbf{static}, and \textbf{conditional} conflicts over temporal validity, factual correctness, and contextual applicability. It simulates controlled long-horizon histories from structured user profiles, introduces cross-session conflicts, and injects semantically similar distractors to create competition among memory candidates. The resulting multi-session dialogue benchmark supports black-box evaluation of final answers and white-box analysis of supporting-memory retrieval and ranking. Experiments on six representative long-term memory systems show uneven strengths across conflict types, with answer correctness often diverging from memory retrieval and ranking. Sensitivity analyses reveal that longer histories, distractors, implicit queries, and larger conflict distances degrade performance. Diagnostics show failures from missing supporting memories and ineffective use of retrieved memories. Collectively, MemConflict advances principled long-term memory governance through retrieval-aware, conflict-aware reliability assessment. Source code and dataset are available at GitHub\footnote{https://github.com/TaoZhen1110/MemConflict}.
\end{abstract}



\received{20 February 2007}
\received[revised]{12 March 2009}
\received[accepted]{5 June 2009}

\maketitle

\section{Introduction}\label{sec1}

Conversational agents based on large language models (LLMs) are increasingly expected to support users in persistent multi-session interactions rather than in isolated conversations \cite{park2023generative, wang2024survey}. In such settings, effective assistance depends not only on understanding the current input, but also on storing, retrieving, and applying user-specific information accumulated from prior interactions, such as personal facts, preferences, ongoing situations, and social relations \cite{zhong2024memorybank}. As a result, long-term memory has become a key component of personalized conversational systems \cite{zhang2025survey}. Rather than merely extending the context window, it is increasingly regarded as a dedicated memory component for maintaining persistent user information over time \cite{packer2023memgpt}.

However, building such systems reliably is challenging because user-specific information is not static: it may evolve over time, coexist with contradictory mentions, or apply only under particular conditions. A memory system may therefore contain earlier states, later updates, contradictory statements, condition-specific preferences, and semantically similar non-target details at the same time \cite{maharana2024evaluating, wu2024longmemeval}. The core challenge is not simply to recall past information, but to retrieve and rank memory items that are valid for the current query. In particular, systems must identify temporally valid states, preserve factually correct information despite contradictions, select contextually applicable preferences, and resist interference from competing alternatives \cite{amiraz2025distracting, xu2024knowledge}. Long-term memory is therefore not merely a storage problem; it also requires validity assessment, memory selection, and conflict-aware ranking \cite{mei2026according}. If these distinctions are not handled reliably, unresolved conflicts can accumulate and undermine the reliability of personalized memory. This motivates evaluation frameworks that assess whether long-term memory systems can retrieve and rank valid memory items under memory conflicts.

Existing evaluations still provide only a partial view of this problem. Prior studies on conflicting information have largely focused on knowledge conflicts in LLMs, retrieval-augmented generation, and open-domain question answering, while conflicts in persistent, user-specific memory remain much less explored \cite{neeman2023disentqa, wang2024memoryllm}. Even within long-term memory evaluation, existing work lacks a diagnostic framework for characterizing different forms of memory conflict. Many studies focus primarily on memory updating, which captures temporal change but leaves other validity challenges underexamined \cite{maharana2024evaluating, chen2025halumem, wu2024longmemeval}. Current evaluations therefore rarely distinguish among several key abilities: identifying temporally valid memory when earlier and later states coexist, preserving factually correct information in the presence of contradictory mentions, and selecting condition-appropriate memory when applicability depends on context \cite{hu2026evermembench, myakala2026beliefshift}. They also provide limited insight into how semantically similar distractors interfere with the retrieval and ranking of valid memory items. More importantly, most evaluations remain answer- or outcome-based, making it difficult to determine whether a failure stems from missing valid memory items, poorly ranked supporting memories, or ineffective use of retrieved memories during response generation \cite{shen2026mem2actbench, chen2025halumem}.

\begin{figure}[t]
    \centering
    \includegraphics[width=\textwidth]{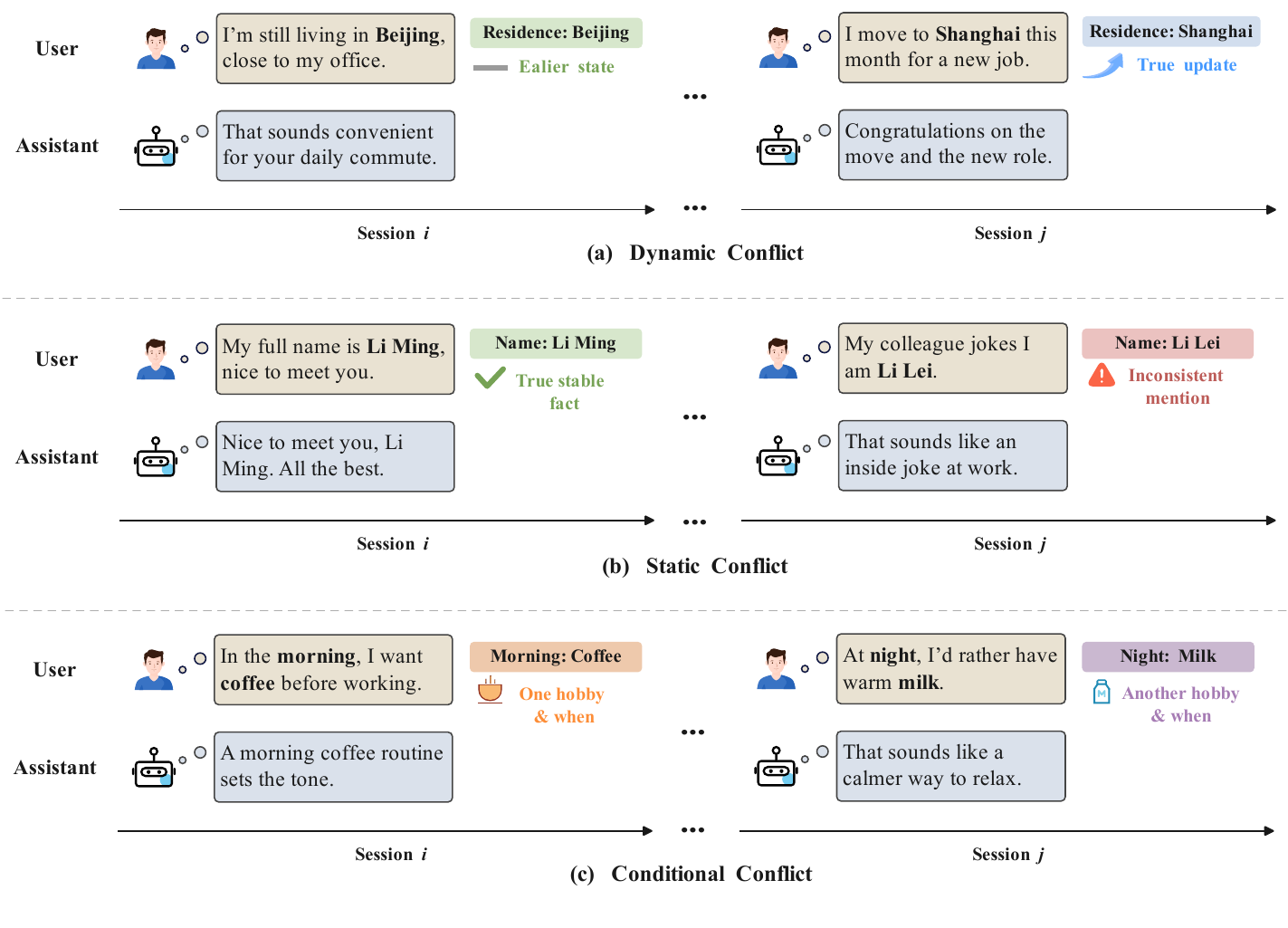}
    \caption{Illustration of three memory conflict types in long-term multi-session conversations. Dynamic conflicts arise when a later true update supersedes an earlier state; static conflicts arise when a later false contradiction should not overwrite a stable fact; conditional conflicts arise when multiple memories are valid under different conditions.}
    \label{fig1}
\end{figure}

To address these limitations, we propose \textbf{MemConflict}, a diagnostic framework for assessing long-term memory systems under memory conflicts. MemConflict is informed by the fitness-for-use perspective from information quality theory \cite{strong1997data, wang1996beyond}, under which information is useful only when it is appropriate for the task and context in which it is used. We adapt this perspective to long-term memory by treating a memory candidate as useful for a query only when it is temporally valid, factually correct, and contextually applicable. These dimensions motivate three conflict types that test whether systems can distinguish valid memory candidates from competing ones that are outdated, contradictory, or conditionally inapplicable. \textbf{Dynamic conflicts} test whether systems can identify the temporally valid memory when user states change over time. \textbf{Static conflicts} test whether systems can retain and retrieve stable user information despite contradictory mentions that should not overwrite it. \textbf{Conditional conflicts} test whether systems can apply the memory whose conditions match the current query, rather than a competing memory that is valid only in another context. Figure \ref{fig1} illustrates examples of how these three conflict types arise in long-term conversational histories.

To instantiate these settings, we simulate long-horizon multi-session histories and construct a benchmark in which user information accumulates sequentially and conflicts are introduced across sessions. Each memory system processes the histories session by session, allowing MemConflict to assess how conflicts are handled as they arise during long-term memory accumulation. We further introduce semantically similar distractors to create realistic competition among memory candidates. MemConflict supports both black-box evaluation of final answers and white-box analysis of memory retrieval and ranking. Together, these evaluations provide diagnostic signals about whether errors are associated with missing valid memories, low-ranked supporting memories, or ineffective use of retrieved memories during response generation. We apply MemConflict to a diverse set of long-term memory systems and show that current systems exhibit distinct weaknesses across dynamic, static, and conditional conflicts, indicating that answer-only evaluation can obscure important memory-level differences in long-term memory reliability.

Our main contributions are as follows:
\begin{itemize}
\item[$\bullet$] To the best of our knowledge, MemConflict is among the first diagnostic evaluation frameworks for memory conflicts in long-term memory systems, formalizing dynamic, static, and conditional conflicts along temporal, factual, and contextual validity dimensions.

\item[$\bullet$] We construct a controlled multi-session benchmark that instantiates these conflicts in long-horizon user histories, introducing cross-session conflicts and semantically similar distractors that require systems to retrieve and rank valid memory items.

\item[$\bullet$] We develop a two-level evaluation protocol that combines black-box answer assessment with white-box memory analysis, separating final-answer correctness from retrieval and ranking of the gold memory item.

\item[$\bullet$] We evaluate six representative long-term memory systems with MemConflict and provide diagnostic findings on their strengths and failure patterns across dynamic, static, and conditional conflicts, deriving implications for evidence-aware and conflict-aware memory system design.
\end{itemize}

The remainder of this paper is organized as follows. Section \ref{sec2} reviews related work on long-term memory systems, long-term memory evaluation, and conflicting information. Section \ref{sec3} presents the MemConflict framework, including timeline simulation, conflict construction, dialogue generation, query construction, and the evaluation protocol. Section \ref{sec4} describes the experimental setup and reports the main results, sensitivity analyses, and system-level diagnostics. Section \ref{sec5} concludes the paper with limitations and directions for future work.

\section{Related Work}\label{sec2}

\subsection{Long-Term Memory Systems}

Interactive artificial intelligence (AI) applications increasingly extend beyond single-session conversations \cite{li2025hello}. In personalized assistants, dialogue agents, and domain-specific support systems, effective responses often depend on user-specific information accumulated across prior interactions, including personal facts, preferences, ongoing situations, and social relations \cite{tan2025prospect, zhang2024llm}. This dependence limits approaches that treat each session in isolation or rely solely on the immediate context window \cite{maharana2024evaluating}. Long-term memory systems address this need by storing, organizing, retrieving, and applying user-specific information across extended interactions.

Recent systems increasingly treat long-term memory as an actively managed component rather than a passive record of dialogue history \cite{li2025memos}. Prior interactions may be represented as user profiles, episodic records, summaries, or hierarchical memory stores, allowing information to be retained at different levels of granularity \cite{packer2023memgpt}. Beyond representation, memory management also involves operations such as writing, consolidation, abstraction, retrieval, and updating \cite{xu2025mem, wang2023augmenting}. This shift positions long-term memory as an infrastructure for maintaining persistent user information, rather than as a byproduct of context accumulation.

However, managing long-term memory remains challenging because user information may change, conflict, or depend on context. In sustained interactions, user-related information can evolve over time, coexist with contradictory mentions, or apply only under specific conditions \cite{yuan2025personalized}. A memory store may therefore contain outdated states, recent updates, contradictory statements, condition-specific preferences, generalized summaries, and semantically similar non-target details at the same time \cite{wan2025storybench, ong2025towards}. The challenge is not only to retrieve relevant past content, but also to determine which memory evidence is valid for the current query. When systems fail to make these distinctions reliably, stale, misleading, or misattributed information may be retained or surfaced during later interactions. These issues make memory validity and consistency central challenges for long-term memory systems, beyond storage alone \cite{hu2025memory}.

Existing work has substantially advanced the design of long-term memory systems, including memory representation, organization, retrieval, and updating. However, it provides more limited insight into how these systems behave when multiple plausible memory candidates compete under temporal change, contradiction, or conditional applicability. Our work builds on this line of research by evaluating whether long-term memory systems can maintain, retrieve, and apply valid memory evidence for the current query under dynamic, static, and conditional conflicts.

\subsection{Evaluation of Memory Systems}

Existing evaluations of long-term memory systems have primarily treated memory as a contributor to downstream interaction and reasoning performance. In long-term conversational settings, benchmarks often test whether systems can use prior interactions for question answering, event summarization, multi-session reasoning, knowledge updating, or memory retrieval \cite{maharana2024evaluating, wu2024longmemeval}. Related work on chat assistants and personalized dialogue evaluates whether recalled memories improve response relevance, emotional support, relationship continuity, and implicit user modeling \cite{zhong2024memorybank, li2025hello, tan2025prospect}. These studies establish the usefulness of memory for persistent interaction, but memory is often evaluated through its contribution to final task performance rather than through direct assessment of how memory is stored, retrieved, selected, and applied.

A second line of work evaluates more specific memory capabilities or system components. Recent benchmarks and analyses examine long-term question answering, temporal reasoning, hallucination in memory systems, personalized referential memory, and task-oriented memory utilization across extended interactions \cite{tan2025membench, wu2024longmemeval, chen2025halumem, hu2026evermembench, mei2026according, shen2026mem2actbench}. Some studies further compare systems with and without explicit memory modules or analyze how memory structures affect consistency and adaptation across sessions \cite{li2025hello, tan2025prospect}. This line of work provides finer-grained evidence than overall task success, but it still tends to evaluate bounded capabilities or individual pipeline stages rather than memory behavior when multiple competing candidates coexist in systematically constructed conflict scenarios. Table~\ref{tab1} summarizes representative long-term memory evaluation frameworks along several dimensions, including evaluation granularity, timing, context length, metrics, conflict coverage, and evaluation tasks.

Conflict-related evaluation remains comparatively narrow in existing memory benchmarks \cite{hu2025evaluating}. When memory conflicts are considered, they are often modeled mainly as update-oriented cases, where a later valid state should supersede an earlier outdated state \cite{maharana2024evaluating, wu2024longmemeval, myakala2026beliefshift}. This setting is important because it captures temporal validity, but it represents only one part of memory validity. Long-term memory systems may also encounter contradictory mentions that should not overwrite stable facts, condition-dependent preferences that apply only in specific contexts, and semantically similar memories that act as distractors. Existing evaluations therefore provide limited coverage of conflict types beyond temporal updating and limited insight into how systems retrieve and rank valid memory items among competing memory candidates.

From an evaluation perspective, conflict settings make it important to look beyond final-answer correctness. A system may fail because the valid memory is not preserved or exposed as retrievable evidence, because it retrieves both valid and invalid memory items but ranks outdated, factually incorrect, conditionally inappropriate, or misattributed memories higher, or because it ineffectively uses retrieved memories during response generation. Conversely, a correct final answer does not necessarily show that the system retrieved or highly ranked the supporting memory item. These distinctions motivate evaluation frameworks that combine black-box answer assessment with memory-level analysis of retrieval and ranking. MemConflict follows this direction by comparing long-term memory systems under controlled conflict scenarios involving temporal updates, contradictory mentions, conditional applicability, and semantically similar distractors.

\begin{table}[t]
  \centering
  \caption{Comparison of Representative Long-Term Memory Evaluation Frameworks.}
  \begin{threeparttable}
  \resizebox{\textwidth}{!}{%
    \begin{tabular}{p{8.55em}p{6.85em}p{5.0em}p{6.35em}p{6.85em}p{6.05em}p{6.95em}}
    \toprule
    \textbf{Framework} &
    \textbf{Granularity} &
    \textbf{Timing} &
    \makecell[l]{\textbf{Avg Length}\\\textbf{/ Session}} &
    \textbf{Metrics} &
    \makecell[l]{\textbf{Conflict}\\\textbf{Coverage}} &
    \makecell[l]{\textbf{Evaluation}\\\textbf{Task}} \\
    \midrule
    \midrule

    LoCoMo~\cite{maharana2024evaluating} &
    End-to-end &
    \makecell[l]{After all\\sessions} &
    477 tokens &
    \makecell[l]{F1; Recall@K;\\ROUGE;\\FactScore;\\BLEU} &
    None &
    \makecell[l]{QA; Generation;\\Summarization} \\
    \midrule

    LongMemEval~\cite{wu2024longmemeval} &
    End-to-end &
    \makecell[l]{After all\\sessions} &
    \mbox{$\sim$3K tokens} &
    \makecell[l]{Accuracy;\\Recall@K} &
    Dynamic &
    QA \\
    \midrule

    PrefEval~\cite{zhao2025llms} &
    End-to-end &
    \makecell[l]{After all\\sessions} &
    \mbox{$\sim$3K tokens} &
    \makecell[l]{Accuracy;\\Preference\\score} &
    Dynamic &
    \makecell[l]{Generation;\\Classification} \\
    \midrule

    PersonaMem~\cite{jiang2025know} &
    End-to-end &
    \makecell[l]{After all\\sessions} &
    \mbox{$\sim$6K tokens} &
    Accuracy &
    Dynamic &
    \makecell[l]{Response\\selection} \\
    \midrule

    MemBench~\cite{tan2025membench} &
    Capability-level &
    \makecell[l]{After each\\interaction\\step} &
    10K tokens &
    \makecell[l]{Accuracy;\\Recall;\\Custom\\scores} &
    Dynamic &
    \makecell[l]{QA; Memory\\retrieval} \\
    \midrule

    HaluMem~\cite{chen2025halumem} &
    Operation-level &
    \makecell[l]{After each\\session} &
    \mbox{$\sim$2.3K tokens} &
    \makecell[l]{Accuracy;\\Recall;\\Custom\\scores} &
    Dynamic &
    \makecell[l]{Memory\\extraction;\\QA} \\
    \midrule

    \makecell[l]{MemoryAgent\\Bench~\cite{hu2025evaluating}} &
    Capability-level &
    \makecell[l]{After each\\interaction\\step} &
    \mbox{$\sim$512 tokens} &
    \makecell[l]{Accuracy;\\Recall@5;\\F1} &
    Dynamic &
    \makecell[l]{QA; Memory\\retrieval} \\
    \midrule

    BeliefShift~\cite{myakala2026beliefshift} &
    End-to-end &
    \makecell[l]{After all\\sessions} &
    \mbox{$\sim$2K tokens} &
    \makecell[l]{Custom\\scores} &
    Dynamic &
    \makecell[l]{QA; Classification} \\
    \midrule

    EverMemBench~\cite{hu2026evermembench} &
    End-to-end &
    \makecell[l]{After all\\sessions} &
    \mbox{$\sim$1M tokens} &
    \makecell[l]{Accuracy;\\Custom\\scores} &
    Dynamic &
    QA \\
    \midrule

    Mem2ActBench~\cite{shen2026mem2actbench} &
    End-to-end &
    \makecell[l]{After all\\sessions} &
    \mbox{$\sim$3.2K tokens} &
    \makecell[l]{F1; BLEU-1;\\Tool Accuracy} &
    Dynamic &
    Tool use \\
    \midrule

    \textbf{MemConflict (Ours)} &
    \makecell[l]{\textbf{Answer-level +}\\\textbf{Memory-level}} &
    \makecell[l]{\textbf{After each}\\\textbf{conflict}\\\textbf{bearing}\\\textbf{session}} &
    \textbf{\mbox{$\sim$3.9K tokens}} &
    \makecell[l]{\textbf{AA; CRS;}\\\textbf{UOCS;}\\\textbf{SEH@K; SRS}} &
    \makecell[l]{\textbf{Dynamic;}\\\textbf{Static;}\\\textbf{Conditional}} &
    \makecell[l]{\textbf{QA; Memory}\\\textbf{retrieval}} \\
    \bottomrule
    \end{tabular}%
  }
  \vspace{2pt}
\begin{tablenotes}[flushleft]
\footnotesize
\item \parbox{\textwidth}{\textit{Note.} Conflict coverage refers to whether a framework systematically constructs and evaluates the corresponding type of memory conflict. AA denotes Answer Accuracy, CRS denotes Conflict Recognition Score, UOCS denotes Update Order Consistency Score, SEH@K denotes Support Evidence Hit@K, and SRS denotes Support Rank Score.}
\end{tablenotes}
  \end{threeparttable}
  \label{tab1}
\end{table}

\subsection{Temporal Dynamics and Information Conflicts}

Temporal dynamics are central to long-term memory because the information maintained by such systems is not fixed over time. Across multi-session interactions, user-related states, situations, and preferences may change, become time-sensitive, or require interpretation relative to when they were stated \cite{wang2023causal, tang2025tcgc, wu2024longmemeval}. Recent benchmarks have begun to evaluate these temporal aspects through tasks involving temporal reasoning, knowledge updating, event tracking across sessions, and timeline inference from distributed dialogue evidence \cite{qin2021timedial, ge2025tremu}. System-oriented work has also introduced time-aware summarization and memory organization mechanisms for long-horizon interaction. These studies mainly examine whether systems can reason over temporal information or recover updated facts, while less directly assessing how systems maintain and rank valid memory items when competing candidates coexist over time.

As memory evolves, conflicting information becomes a natural consequence of long-term interaction rather than a separate issue. Recent research on LLMs and retrieval-augmented generation (RAG) systems has examined related forms of conflict, including inconsistencies between retrieved context and parametric knowledge, contradictions across retrieved contexts, and competition among semantically similar but incompatible evidence \cite{longpre2021entity, neeman2023disentqa}. These studies show that conflict resolution cannot be reduced to topical relevance alone: a system may retrieve plausible evidence yet still fail to recognize which evidence is valid for the current query or to prioritize the appropriate information during generation \cite{cuconasu2024power, chen2024benchmarking}. At the same time, most of this work is grounded in open-domain question answering or RAG settings, where the main concern is conflict among documents, retrieved passages, or model knowledge \cite{xu2024knowledge, xiong2025icr, lyu2025crud}. Long-term memory systems face a related but distinct problem, in which conflicts arise from evolving user states, contradictory mentions, condition-dependent preferences, and semantically similar distractors accumulated across sessions.

These lines of work reveal a gap in current memory evaluation. Existing memory benchmarks mainly emphasize recall, reasoning, or downstream usefulness, whereas conflict-oriented studies usually focus on open-domain evidence or interactions between retrieved context and model knowledge. These two lines of work remain only loosely connected for long-term memory systems operating over evolving user information. What is still missing is a diagnostic evaluation framework for assessing whether systems can maintain, retrieve, rank, and apply valid memory across temporal updates, contradictory mentions, condition-dependent applicability, and semantically similar distractors. MemConflict addresses this gap by systematically evaluating long-term memory systems under dynamic, static, and conditional conflicts, while also diagnosing memory retrieval and ranking under competing memory candidates.

\section{The MemConflict Framework}\label{sec3}

\begin{figure}[t]
    \centering
    \includegraphics[width=\textwidth]{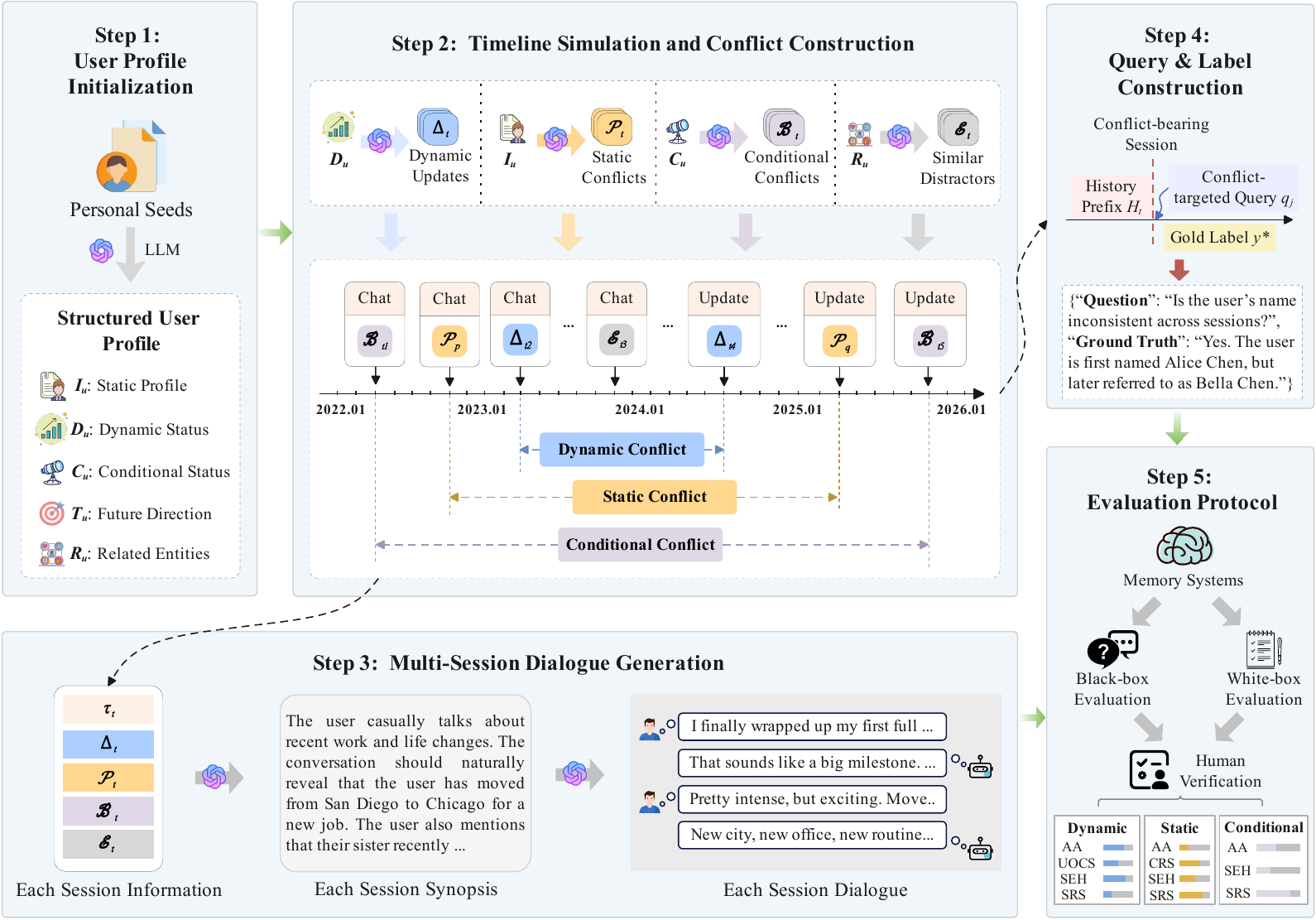}
\caption{Overview of the MemConflict framework. Starting from structured user profiles, the framework simulates session timelines and inserts dynamic updates $\Delta_t$, static information $\mathcal{P}_t$, conditional bindings $\mathcal{B}_t$, and distractors $\mathcal{E}_t$ before generating dialogues, queries, and evaluation labels. Subscripts denote insertion sessions.}
    \label{fig2}
\end{figure}

In this section, we present MemConflict, a framework for evaluating long-term memory systems under memory conflicts. MemConflict assesses whether a system can retrieve and apply memory evidence that is temporally valid, factually correct, and contextually applicable in long-horizon, multi-session interactions. The framework combines conflict-oriented benchmark construction with a diagnostic evaluation protocol: starting from structured user profiles and related entities, it simulates evolving cross-session interactions, instantiates dynamic, static, and conditional conflicts, and converts them into benchmark dialogues and conflict-targeted queries. The resulting benchmark supports both black-box response evaluation and white-box memory retrieval and ranking analysis. We first formulate the evaluation problem and then describe the five main components of MemConflict: user profile initialization, timeline simulation and conflict construction, dialogue generation, query construction, and the evaluation protocol. Figure~\ref{fig2} provides an overview of the framework. Representative prompt templates for the LLM-assisted construction and evaluation steps are provided in Appendix~\ref{sec_a1}.

\subsection{Problem Formulation}\label{sec3.1}

In long-horizon, cross-session interactions, the same user attribute may be associated with multiple candidate memory items over time. These candidates may result from true state updates, contradictory mentions, condition-dependent values, or semantically similar non-target information distributed across the interaction history. A long-term memory system must therefore not only recall previously mentioned content, but also identify which memory item is valid for the current query. We formulate this challenge as a validity determination problem: given a query and a temporally ordered interaction history, the system must return the memory content that is valid for that query.

Following the fitness-for-use perspective \cite{strong1997data, wang1996beyond}, MemConflict defines memory validity along three complementary dimensions: temporal validity, factual correctness, and contextual applicability. A memory item is temporally valid if it matches the true user state at query time, factually correct if it remains the benchmark-defined true value despite contradictory mentions, and contextually applicable if it is appropriate under the condition specified by the query. Dynamic conflicts arise when earlier and later states coexist and validity depends on which state is temporally valid after the current session. Static conflicts arise when contradictory mentions challenge factual correctness without representing true state updates. Conditional conflicts arise when multiple values appear plausible, but only the value-condition association matching the query context is applicable.

We formalize this setting as follows. Let $H_t = (s_1, s_2, \dots, s_t)$ denote the temporally ordered interaction history available to the memory system after session $s_t$ has been ingested. In MemConflict, each query is issued immediately after the system ingests a session that introduces information conflicting or competing with previously mentioned memory candidates. Let $q_t$ denote a query issued after $s_t$ about a target attribute $\alpha$, and let $\mathcal{C}(\alpha, H_t)$ denote the set of candidate memory items for $\alpha$ available in $H_t$. A memory conflict arises when multiple candidates in $\mathcal{C}(\alpha, H_t)$ compete as possible answers for the same query, but differ in temporal validity, factual correctness, or contextual applicability. MemConflict treats each query as grounded in the history prefix available at query time, rather than in the full interaction history. The gold candidate for query $q_t$ is defined as:
\begin{equation}
    v^*(\alpha, q_t \mid H_t) =
    \operatorname*{arg\,max}_{v \in \mathcal{C}(\alpha, H_t)}
    \mathcal{F}(v \mid \alpha, H_t, q_t),
\end{equation}
where $\mathcal{F}(v \mid \alpha, H_t, q_t)$ denotes the benchmark-defined, query-conditioned validity of candidate $v$ according to temporal validity, factual correctness, and contextual applicability. The benchmark construction guarantees a unique maximizer for each query. Under this formulation, dynamic, static, and conditional conflict instances respectively evaluate temporal update handling, factual preservation under contradictory mentions, and recovery of the value-condition association applicable to the query. The task is therefore not simply to recover any previously mentioned content, but to identify the memory content that is valid for the current query while remaining sensitive to competing candidates in the available history.

\subsection{User Profile Initialization}\label{sec3.2}

Rather than constructing conflicts as isolated attribute edits, MemConflict begins by initializing a coherent user context for each benchmark instance. Each virtual user is initialized with a structured profile containing dynamic states, invariant background information, condition-dependent preferences, personality traits and life goals, and related entities. This profile provides the basis for generating subsequent events, state transitions, conflict instances, and distractors in a natural and interpretable manner.

To improve realism and diversity, MemConflict initializes each user from a persona seed $z_u$ sampled from Persona Hub\footnote{https://github.com/tencent-ailab/persona-hub}. Conditioned on $z_u$, an LLM generates the structured profile of user $u$ according to predefined attribute schemas:
\begin{equation}
    U_u = \mathrm{LLM}(z_u) = (I_u, D_u, C_u, T_u, R_u),
\end{equation}
where $U_u$ denotes the structured profile of user $u$, $I_u$ denotes invariant profile attributes, $D_u$ denotes dynamic state attributes, $C_u$ denotes conditional preference attributes, $T_u$ denotes personality traits and life goals, and $R_u$ denotes related-entity information. Dynamic state attributes include current residence, marital status, child status, employment status, health state, social state, and career direction, and serve as the primary source of temporal state transitions. Invariant profile attributes include basic personal information, birthplace, family background, and educational background, and remain stable throughout the evaluation horizon. Conditional preference attributes encode stable but condition-dependent preference patterns and provide the basis for conditional conflicts. Personality traits and life goals constrain the direction, pace, and plausibility of later state changes and event development. Related-entity information describes other entities in the user's environment and supports the construction of semantically similar distractors.

\subsection{Timeline Simulation and Conflict Construction}\label{sec3.3}

Building on the initialized user profiles, MemConflict constructs long-horizon, cross-session interaction histories in which user-related information accumulates, changes, and is revisited across sessions. The goal of this stage is not simply to generate long interaction traces, but to operationalize temporal validity, factual correctness, and contextual applicability as concrete benchmark instances embedded in coherent user trajectories. To this end, MemConflict first simulates plausible user evolution over time and then instantiates dynamic, static, and conditional conflicts within the resulting timelines. It further introduces semantically similar distractors to create realistic competition among memory candidates. The following subsections describe the timeline simulation procedure and the construction of conflict instances and distractors.

\subsubsection{\textbf{Timeline Simulation}}\label{sec3.3.1}

For each initialized user profile $U_u = (I_u, D_u, C_u, T_u, R_u)$, MemConflict simulates a temporally ordered sequence of sessions that serves as the interaction timeline for user $u$. This stage is designed to approximate realistic long-horizon user interaction, in which user-related information is gradually introduced, updated, and revisited across sessions. The resulting multi-session history provides a coherent basis for subsequent conflict instantiation.

The timeline spans January 2022 to December 2025, with observed sessions organized at monthly granularity. We begin the simulation in 2022 to align the benchmark horizon with the recent rise of LLM-based interactive applications, while the four-year span provides sufficient room for state evolution and conflict accumulation. For user $u$, the observed interaction timeline is represented as:
\begin{equation}
    \mathcal{T}_u = (s_1, s_2, \dots, s_T),
\end{equation}
where each session $s_t$ corresponds to the $t$-th temporal slot and is associated with a session type $\tau_t \in \{\text{chat}, \text{update}\}$. Chat sessions provide non-update interaction opportunities, whereas update sessions are reserved for genuine changes in the user's dynamic state. These session slots serve as temporal positions into which initial mentions, state updates, conflict values, and distractors are placed in subsequent construction steps. Under this construction, the query-time history prefix $H_t$ defined in Section \ref{sec3.1} corresponds to the first $t$ observed sessions in $\mathcal{T}_u$.

The initial dynamic attributes $D_u$ defined in Section \ref{sec3.2} provide the user's starting ground-truth state. To avoid overly mechanical information exposure, mentions of these initial attributes are distributed across early sessions rather than revealed in a single initial session. The dynamic state trajectory induced by update sessions is constructed next.

\subsubsection{\textbf{Dynamic Conflict Construction via State Transitions}}

Starting from the session slots constructed in Section \ref{sec3.3.1}, MemConflict simulates how the user's dynamic state evolves over time. The goal is to create realistic state changes rather than arbitrary attribute edits. In each update session, only a small subset of dynamic attributes is allowed to change, reflecting the fact that only a limited number of aspects of a user's life typically change at any given time. Each accepted state change creates a dynamic conflict when the previous value and the updated value both remain available in the interaction history, but only the updated value is valid after the update session.

Let $\mathcal{A}_D$ denote the set of dynamic attributes represented in the initialized user state $D_u$, including current residence, marital status, child status, employment status, health state, social state, and career direction. Let $D_t$ denote the ground-truth dynamic state after session $s_t$. We initialize the state trajectory with $D_0 = D_u$. For a chat session, no dynamic update is introduced, and we set $D_t = D_{t-1}$. For an update session, MemConflict selects an update set $A_t \subseteq \mathcal{A}_D$ with $|A_t| \in \{1,2\}$.

The update set $A_t$ is selected under feasibility, cooldown, and persona-consistency constraints. Feasibility requires that an attribute can be updated only when its transition preconditions are satisfied. Cooldown prevents immediate reversals or repeated updates in adjacent sessions unless a sufficient trigger is available. Persona consistency biases updates toward directions compatible with the user's personality traits and life goals encoded in $T_u$. Attribute-specific update weights are also used, assigning higher update probabilities to health state, social state, and employment status than to current residence, marital status, child status, and career direction. These constraints and weights are used to better approximate real-life patterns of user state changes.

For the selected update set $A_t$, MemConflict uses an LLM to propose candidate new values:
\begin{equation}
    \{\hat{v}_{t,d}: d \in A_t\} = \mathrm{LLM}(A_t, D_{t-1}, T_u),
\end{equation}
where $\hat{v}_{t,d}$ denotes the candidate new value proposed for attribute $d$ in session $s_t$. MemConflict retains only proposed values that pass attribute-specific validation and differ from their previous values. The accepted changes for session $s_t$ are defined as:
\begin{equation}
    \Delta_t =
    \{(d, \hat{v}_{t,d}) : d \in A_t,\ 
    V_d(D_{t-1}, \hat{v}_{t,d})=1
    \land \hat{v}_{t,d} \neq D_{t-1}[d]\},
\end{equation}
where $V_d(\cdot)$ denotes the attribute-specific validation rule, which enforces structural constraints such as marital-status progressions, required preconditions for child status changes, and longer cooldown intervals for structural changes such as residence or career direction updates. For each $(d,\hat{v}_{t,d}) \in \Delta_t$, the ground-truth state is updated with $D_t[d]=\hat{v}_{t,d}$; all other attributes inherit their previous values from $D_{t-1}$. The accepted changes in $\Delta_t$ are inserted into session $s_t$ as explicit update mentions, while unchanged attributes remain part of the ground-truth state trajectory but are not re-mentioned in that session. For chat sessions, no candidate values are generated and $\Delta_t=\emptyset$.

\subsubsection{\textbf{Static Conflict Construction}}

Static conflicts test whether a memory system can preserve the true value when a later contradiction does not reflect an actual state update. In MemConflict, static conflicts are constructed only over invariant profile attributes, which remain unchanged throughout the evaluation horizon. Let $\mathcal{A}_I$ denote the set of invariant attributes represented in the initialized profile $I_u$, including basic personal information, birthplace, family background, and educational background. For any target attribute $b \in \mathcal{A}_I$, the gold value is fixed by the initialized profile and does not change over time.

To instantiate a static conflict for user $u$, MemConflict selects a target attribute $b \in \mathcal{A}_I$ and two observed sessions $s_p, s_q \in \mathcal{T}_u$ such that $p < q \leq T$. At session $s_p$, the true value $v^* = I_u[b]$ is inserted as an explicit mention of the target user's invariant attribute. At a later session $s_q$, MemConflict uses an LLM to generate a plausible but contradictory value $v^-$ for the same attribute and inserts it as a contradictory mention, where $v^- \neq v^*$. Unlike dynamic conflicts, the later mention at $s_q$ is explicitly treated as false information rather than as a valid state update. To isolate the contradiction, the target attribute $b$ is not mentioned in the intermediate sessions $s_{p+1}, \dots, s_{q-1}$. We use $\mathcal{P}_t$ to denote the static information inserted into session $s_t$, with $\mathcal{P}_p=\{(b,v^*)\}$ for the true mention and $\mathcal{P}_q=\{(b,v^-)\}$ for the contradictory mention. This construction creates a factual conflict in which the target user's true invariant value and a later false mention coexist in the available history.

\subsubsection{\textbf{Conditional Conflict Construction}}

Conditional conflicts test whether a memory system can preserve the association between a preference value and the condition under which it applies. In MemConflict, such conflicts are constructed over conditional preference attributes, which encode stable but context-dependent preference patterns rather than unconditional values. Let $\mathcal{A}_C$ denote the set of conditional preference attributes in the initialized user profile, and let $C_u[r]$ denote the set of condition-value pairs associated with attribute $r \in \mathcal{A}_C$. A conditional conflict arises when multiple values for the same attribute are all valid under their respective conditions, but each value is applicable only under its associated condition. For example, a user may prefer coffee in the morning and milk in the evening; both preferences are valid, but their applicability depends on the time of day.

To instantiate a conditional conflict for user $u$, MemConflict selects a target attribute $r \in \mathcal{A}_C$ and a set of condition-value pairs $\{(c_\ell, v_\ell)\}_{\ell=1}^{L} \subseteq C_u[r]$, where $L \geq 2$ and the selected conditions are distinct. These pairs are introduced into the interaction history at different observed sessions in $\mathcal{T}_u$. Unlike static or dynamic conflicts, the values are factually correct and do not represent temporal updates of one another. The apparent conflict arises because the values are associated with different contextual conditions, and only one is applicable to a given query. To isolate the conditional contrast, the target attribute $r$ is not re-mentioned between the sessions in which these condition-value pairs are introduced. We use $\mathcal{B}_t$ to denote the conditional bindings inserted into session $s_t$; when a pair $(c_\ell, v_\ell)$ is introduced in session $s_t$, $(r, c_\ell, v_\ell)$ is added to $\mathcal{B}_t$. This construction creates contextual competition among multiple user-consistent but condition-specific preferences.

\subsubsection{\textbf{Distractor Injection}}

Distractor injection tests whether a memory system can distinguish target-user memories from semantically similar but non-target information. In MemConflict, distractors are selected from the related-entity information $R_u$ in the initialized user profile. A distractor is factually correct for another entity in the user's environment, but it is not a valid memory for the target user $u$. In this way, distractors introduce realistic competition among memory candidates.

For each constructed conflict instance involving target attribute $\alpha$, MemConflict examines $R_u$ for related-entity information that is semantically similar to the target information involved in the conflict. If such information is available, it is selected as a distractor and inserted into an intermediate session within the conflict span, i.e., between the sessions where the target conflict information is introduced. In static conflicts, the selected distractors resemble invariant facts of the target user; in dynamic conflicts, they resemble user state values; and in conditional conflicts, they resemble condition-dependent preferences. In all cases, the distractor remains attributable to a related entity rather than to the target user. Let $t_{\mathrm{first}}$ and $t_{\mathrm{last}}$ denote the first and last sessions in which the target conflict information is introduced. We use $\mathcal{E}_t$ to denote the distractor information inserted into an intermediate session $s_t$ within this span:
\begin{equation}
    \mathcal{E}_t = \{(e, \alpha, v_e) : e \in R_u,\ e \neq u,\ t_{\mathrm{first}} < t < t_{\mathrm{last}}\},
\end{equation}
where $e$ denotes a related entity represented in $R_u$ and $v_e$ denotes a semantically similar non-target value for the target attribute $\alpha$. This design increases retrieval and ranking difficulty while leaving the correct answer unchanged.

\subsection{Multi-Session Dialogue Generation}

Given the simulated timeline $\mathcal{T}_u$ and the session-level information constructed in Section \ref{sec3.3}, MemConflict converts each session slot into a natural multi-turn dialogue. For session $s_t$, the information to be verbalized may include dynamic updates $\Delta_t$, static information $\mathcal{P}_t$, conditional bindings $\mathcal{B}_t$, and distractor information $\mathcal{E}_t$. The goal of dialogue generation is to express these items naturally in conversation while preserving the coherence of the user's long-horizon interaction history. MemConflict uses a two-stage process: synopsis generation followed by dialogue realization.

In the first stage, MemConflict constructs a session synopsis $z_t$ that specifies the communicative focus of session $s_t$ and organizes the information to be expressed. The synopsis is generated as:
\begin{equation}
    z_t = \mathrm{LLM}(\tau_t, \Delta_t, \mathcal{P}_t, \mathcal{B}_t, \mathcal{E}_t),
\end{equation}
where $\tau_t$ is the session type, and $\Delta_t$, $\mathcal{P}_t$, $\mathcal{B}_t$, and $\mathcal{E}_t$ denote the dynamic updates, static information, conditional bindings, and distractor information assigned to session $s_t$, respectively. Some of these information sets may be empty when the corresponding information type is not assigned to session $s_t$. For update sessions, the synopsis foregrounds the state changes in $\Delta_t$. For chat sessions, it is anchored in an everyday conversational topic while still providing a natural opportunity to express any assigned static, conditional, or distractor information. The synopsis therefore aligns session type, conversational topic, and required memory content before full dialogue generation.

In the second stage, MemConflict expands the synopsis into a multi-turn dialogue:
\begin{equation}
    G_t = \mathrm{LLM}(z_t, \Delta_t, \mathcal{P}_t, \mathcal{B}_t, \mathcal{E}_t),
\end{equation}
where $G_t = (g_{t,1}, g_{t,2}, \dots, g_{t,N_t})$ and $N_t$ is the number of dialogue turns in session $t$. The dialogue must satisfy two constraints. First, it must remain coherent with the session synopsis and conversational topic. Second, every non-empty information item in $\Delta_t$, $\mathcal{P}_t$, $\mathcal{B}_t$, and $\mathcal{E}_t$ must be explicitly realized in the dialogue, ensuring that state updates, static mentions, condition-value bindings, and distractors are observable from the interaction history. Rather than inserting these items as lists, MemConflict distributes them across turns and integrates them into the conversational flow.

After generation, human experts inspect the dialogues for naturalness, coherence, and information coverage. They verify that all required items are explicitly expressed and correctly attributed to the target user or related entities. Dialogues that fail these checks are revised or regenerated before inclusion in the benchmark.

\subsection{Query and Label Construction}

MemConflict constructs evaluation queries only after a conflict instance becomes evaluable, rather than at every session in the simulated timeline. A conflict becomes evaluable once all information required to define the gold answer has appeared in the available history. For each evaluable instance, MemConflict issues a query immediately after the corresponding session $s_t$ has been ingested and uses the history prefix $H_t$ as the memory available to the system. Non-conflict sessions do not directly generate queries, but they remain part of $H_t$ and may provide supporting context or distractor information.

Each query is associated with a conflict instance, a target attribute $\alpha$, and a conflict type $\kappa \in \{\text{dynamic}, \text{static}, \text{conditional}\}$. For static conflicts, the query is issued after the later contradictory mention has been introduced and asks for the true value of the invariant attribute. For dynamic conflicts, the query is issued after a true state update has occurred and asks for the temporally valid value of the dynamic attribute at query time. For conditional conflicts, the query specifies a preference value and asks for the condition under which that value applies. This design tests whether the system can recover the correct value-condition association rather than relying on the most recent or most salient preference mention.

Let $Q_u$ denote the set of evaluation queries generated for user $u$. Each query $q_j \in Q_u$ is associated with an evaluable session index $t_j$, a target attribute $\alpha_j$, a conflict type $\kappa_j$, and a history prefix $H_{t_j}$. For static and dynamic conflicts, the gold answer is the valid memory value:
\begin{equation}
    y^*(q_j \mid H_{t_j}) = v^*(\alpha_j, q_j \mid H_{t_j}).
\end{equation}
For conditional conflicts, suppose the queried value is $\nu(q_j)$ and the target user's conditional profile contains an introduced pair $(c_\ell,\nu(q_j)) \in C_u[\alpha_j]$. The construction ensures that the queried value maps to a unique condition, and the gold answer is the corresponding condition:
\begin{equation}
    y^*(q_j \mid H_{t_j}) = c_\ell.
\end{equation}
For selected instances, MemConflict also includes diagnostic queries that ask whether competing candidates or update relations are present in the available history. This construction converts conflict-bearing histories into testable evaluation instances while preserving the distinction among temporal validity, factual correctness, and contextual applicability.

Algorithm~\ref{alg1} summarizes the construction pipeline described above. It shows how MemConflict initializes user profiles, simulates session timelines, injects dynamic updates, static mentions, conditional bindings, and distractors into sessions, generates multi-session dialogues, and finally constructs evaluation queries and gold labels.

\begin{algorithm}[t]
\caption{MemConflict Benchmark Construction}
\label{alg1}
\begin{algorithmic}[1]
\Require Persona seeds $\{z_u\}$, construction configuration
\Ensure Multi-session dialogues $\mathcal{G}$, evaluation queries $Q$, gold labels $Y$
\For{each user $u$}
    \State Generate initialized profile $U_u=(I_u,D_u,C_u,T_u,R_u)$ from seed $z_u$
    \State Simulate session timeline $\mathcal{T}_u=(s_1,\dots,s_T)$
    \State Insert initial profile mentions into early sessions
    \State Initialize dynamic state $D_0 \gets D_u$
    \For{each session slot $s_t \in \mathcal{T}_u$}
        \If{$s_t$ is an update session}
            \State Select update attributes $A_t \subseteq \mathcal{A}_D$
            \State Generate and validate accepted updates $\Delta_t$
            \State Insert $\Delta_t$ into session $s_t$ and update dynamic state
        \EndIf
    \EndFor
    \State Insert static conflict mentions $\mathcal{P}_t$ into selected sessions
    \State Insert conditional bindings $\mathcal{B}_t$ into selected sessions
    \State Inject related-entity distractors $\mathcal{E}_t$ within conflict spans
    \For{each session $s_t \in \mathcal{T}_u$}
        \State Generate synopsis $z_t$ from $\tau_t,\Delta_t,\mathcal{P}_t,\mathcal{B}_t,\mathcal{E}_t$
        \State Generate dialogue $G_t$ from $z_t$ and inserted information
    \EndFor
    \State Construct evaluation queries and gold labels from evaluable conflict instances
\EndFor
\State \Return dialogues $\mathcal{G}$, queries $Q$, labels $Y$
\end{algorithmic}
\end{algorithm}

\subsection{Evaluation Protocol}\label{sec3.6}

MemConflict evaluates each query at two levels. Black-box evaluation measures whether the system returns the correct final answer, while white-box evaluation examines whether the retrieved memory items contain the memory supporting the correct answer and how highly it is ranked. Let $q_i$ denote the $i$-th query, $\hat{y}_i$ the system answer, $y_i^*$ the gold answer, $R_i^K$ the top-$K$ retrieved memory items, and $m_i^*$ the gold memory item whose content is consistent with $y_i^*$. Let $N$ be the number of queries in the corresponding subset. All metrics are reported separately for dynamic, static, and conditional conflict subsets.

For black-box evaluation, we use Answer Accuracy (AA):
\begin{equation}
    \text{AA} = \frac{1}{N} \sum_{i=1}^N \mathbf{1}(\hat{y}_i = y_i^*),
\end{equation}
where, for dynamic and static conflicts, $y_i^*$ is the valid memory value; for conditional conflicts, it is the condition associated with the queried preference value.

For white-box evaluation, we first check whether the top-$K$ retrieved memory items contain the memory item consistent with the gold answer. Support Evidence Hit@K (SEH@K) is defined as:
\begin{equation}
    \text{SEH@K} = \frac{1}{N} \sum_{i=1}^N \mathbf{1}(m_i^* \in R_i^K).
\end{equation}
If the supporting memory item is retrieved, we further evaluate how highly it is ranked in $R_i^K$. Support Rank Score (SRS) assigns a logarithmically discounted score according to its rank, and assigns 0 when it is not retrieved:
\begin{equation}
    \text{SRS} = \frac{1}{N} \sum_{i=1}^N
    \begin{cases}
    \frac{1}{\log_2(\mathrm{rank}(m_i^*) + 1)}, & \text{if } m_i^* \in R_i^K, \\
    0, & \text{otherwise}.
    \end{cases}
\end{equation}

MemConflict further includes lightweight conflict-type diagnostics when the final answer alone is insufficient. For dynamic conflicts, Update Order Consistency Score (UOCS) measures whether the system recognizes that a true update has occurred and preserves the temporal order between earlier and later states:
\begin{equation}
    \text{UOCS} = \frac{1}{N} \sum_{i=1}^N \mathbf{1}(u_i = 1 \land o_i = 1),
\end{equation}
where $u_i$ and $o_i$ indicate correct update recognition and temporal ordering, respectively. For static conflicts, Conflict Recognition Score (CRS) measures whether the system identifies the presence of contradictory information:
\begin{equation}
    \text{CRS} = \frac{1}{N} \sum_{i=1}^N \mathbf{1}(c_i = 1),
\end{equation}
where $c_i$ indicates correct recognition of incompatible candidates in the available history. No additional diagnostic metric is introduced for conditional conflicts, because the corresponding queries directly evaluate recovery and ranking of the correct value-condition association. Together, these metrics assess final-answer correctness, memory retrieval, and memory ranking under temporal, factual, and contextual memory conflicts.

All answer and memory-item judgments are produced through LLM-assisted matching followed by human verification. Human annotators review the LLM judgments, especially ambiguous matches, paraphrases, and attribution cases, and the final metrics are computed from the verified judgments.

\subsection{Dataset Statistics}\label{sec3.7}

Table~\ref{tab:mceval_stats} reports average per-instance statistics of the MemConflict dataset. We constructed 12 benchmark instances for the experiments. Each instance corresponds to one virtual user and contains a long-horizon multi-session dialogue history, conflict-bearing sessions, related-entity distractors, and evaluation queries. On average, each instance contains 52.33 sessions, 2,349.17 dialogue turns, and 203,910.83 tokens, with 124.33 evaluation queries. These statistics show that MemConflict evaluates memory behavior under extended multi-session histories rather than short isolated conversations.

The distribution of conflict queries is intentionally not uniform. Dynamic conflicts account for the largest portion because user states naturally evolve in long-term interactions, whereas static contradictions and conditional preference conflicts are introduced more conservatively to preserve clear factual and contextual labels. The conflict-distance ranges further vary the separation between competing memory candidates, testing whether systems can connect related information across both nearby and distant sessions. Distractor entries from related entities are also injected to create semantically similar non-target memories, making memory retrieval and ranking closer to realistic personalized memory use. Because the three conflict types contain different numbers of queries, we report metrics separately by conflict type and compute overall averages as macro-averages over conflict types, so that aggregate scores are not dominated by the more frequent dynamic conflicts.

\begin{table}[t]
\centering
\caption{Statistical Overview of the MemConflict Dataset.}
\label{tab:mceval_stats}
\small
\renewcommand{\arraystretch}{1.08}
\begin{tabular}{p{0.58\textwidth}r}
\toprule
\textbf{Metric} & \textbf{Average per Instance} \\
\midrule

\multicolumn{2}{l}{\textbf{Interaction Statistics}} \\
Average Session Number & 52.33 \\
Average Dialogue Turns & 2,349.17 \\
Average Context Length (tokens) & 203,910.83 \\
Average Dialogue Turns per Session & 44.89 \\
Average Tokens per Session & 3,896.39 \\

\midrule
\multicolumn{2}{l}{\textbf{Conflict Statistics}} \\
Dynamic Conflict Distance & 5--25 sessions \\
Static Conflict Distance & 10--45 sessions \\
Conditional Conflict Distance & 9--49 sessions \\

\midrule
\multicolumn{2}{l}{\textbf{Distractor Statistics}} \\
Average Distractor Entries & 32.83 \\
Average Distractor Entries per Session & 0.63 \\

\midrule
\multicolumn{2}{l}{\textbf{Question Statistics}} \\
Average Questions per Instance & 124.33 \\
Dynamic Conflict Questions & 90.82 \\
Static Conflict Questions & 16.65 \\
Conditional Conflict Questions & 16.86 \\

\bottomrule
\end{tabular}
\end{table}

\section{Experiments}\label{sec4}

In this section, we evaluate representative long-term memory systems on MemConflict to assess both system performance and the diagnostic value of the framework. We compare system behavior across dynamic, static, and conditional conflicts using black-box answer evaluation and white-box memory retrieval and ranking analysis, examining whether systems not only answer correctly but also retrieve and rank the memory item supporting the correct answer. We further analyze the effects of controlled benchmark factors and diagnose efficiency and failure patterns. These experiments show how MemConflict reveals weaknesses that answer-only evaluation may overlook and what they imply for evidence-aware and conflict-aware memory system design.

\subsection{Evaluated Memory Systems}

We evaluate six representative long-term memory systems: A-Mem \cite{xu2025mem}, LangMem\footnote{https://github.com/langchain-ai/langmem}, Letta\footnote{https://github.com/letta-ai/letta}, MemOS \cite{li2025memos}, Mem0 \cite{chhikara2025mem0}, and Memobase\footnote{https://github.com/memodb-io/memobase}. These systems were selected to cover a range of publicly available memory designs for LLM-based agents and applications, including agentic memory organization, toolkit-based memory management, persistent agent memory, explicit memory lifecycle control, lightweight memory layers, and profile-centric user memory. Evaluating them under a common benchmark allows us to examine how different memory designs behave when correct performance depends on temporal validity, factual correctness, and contextual applicability.

\begin{itemize}
\item \textbf{A-Mem}. A-Mem is an agentic memory system that organizes memories as linked notes rather than storing them as a flat collection, allowing us to examine how explicitly structured memory behaves under conflicting information.

\item \textbf{LangMem}. LangMem is a toolkit-oriented long-term memory framework that supports memory extraction, consolidation, and updating for downstream agent workflows, representing practical memory management beyond raw interaction-history retrieval.

\item \textbf{Letta}. Letta is a persistent agent framework in which long-term memory forms part of an agent's ongoing state across sessions through a managed memory hierarchy, making it suitable for evaluating persistent agent memory under distributed competing memories.

\item \textbf{MemOS}. MemOS treats memory as an explicit system resource with dedicated mechanisms for representation, storage, retrieval, and lifecycle management, allowing us to assess explicit memory management under different forms of memory conflict.

\item \textbf{Mem0}. Mem0 is a lightweight persistent memory layer that emphasizes efficient memory extraction and retrieval with minimal integration overhead, representing lightweight memory designs where validity depends on temporal change and conflict resolution.

\item \textbf{Memobase}. Memobase is a profile-centric memory backend that maintains user memory through structured profiles and event timelines, representing profile-based user memory in personalized settings.
\end{itemize}

To ensure comparability, all six systems are integrated into the same MemConflict query and evaluation pipeline. Unless otherwise specified, all systems follow the same query protocol, answer normalization procedure, and white-box memory retrieval interface, while each method is kept as close as possible to its intended memory design and default usage.

\subsection{Experimental Setup}

We evaluate all systems under the same MemConflict query set, answer normalization procedure, and white-box memory retrieval and ranking protocol. The experiments vary the following factors:

\begin{itemize}
\item \textbf{Conflict Types}. All systems are evaluated on dynamic, static, and conditional conflicts, corresponding to temporal validity, factual correctness, and contextual applicability, respectively.

\item \textbf{Evaluation Perspectives}. Black-box evaluation measures final-answer correctness, while white-box evaluation examines whether retrieved memory items contain the gold memory item and how highly it is ranked.

\item \textbf{Retrieval Depth}. The main white-box evaluation uses \(K=3\). We further examine retrieval sensitivity with \(K \in \{2,5\}\).

\item \textbf{Dialogue Length}. Dialogue length controls the amount of surrounding conversational context between target information and the query. We compare medium and long dialogue settings.

\item \textbf{Distractor Injection}. We compare settings with and without semantically similar but non-target information from related entities.

\item \textbf{Query Formulation}. We evaluate both direct queries and implicit queries that require less explicit grounding of the target memory.

\item \textbf{Conflict Distance}. Conflict distance measures how far apart competing memory candidates are distributed across sessions. We evaluate distance variants by changing the number of intervening sessions between competing memory candidates.

\item \textbf{Metrics}. We report Answer Accuracy (AA) for all conflict types. Update Order Consistency Score (UOCS) is reported for dynamic conflicts, and Conflict Recognition Score (CRS) for static conflicts. Support Evidence Hit@K (SEH@K) and Support Rank Score (SRS) are used for white-box analysis across all conflict types, following the definitions in Section \ref{sec3.6}.
\end{itemize}

\begin{figure}[t]
\centering
\includegraphics[width=13cm]{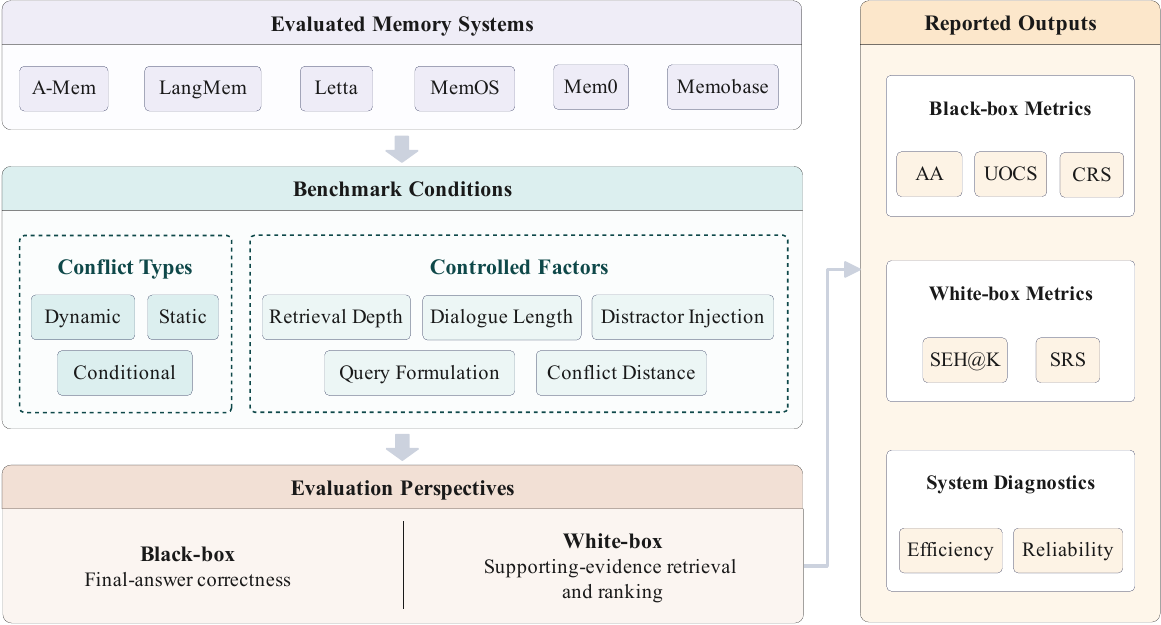}
\caption{Overview of evaluation dimensions and reported outputs in MemConflict.}
\label{fig3}
\end{figure}

Figure~\ref{fig3} summarizes the evaluation dimensions and reported outputs used in MemConflict. Unless otherwise specified, the main results use the default configuration with direct queries, medium dialogue length, distractor injection, medium conflict distance, and \(K=3\) for white-box analysis. Each sensitivity analysis varies one factor while keeping the others fixed. For black-box evaluation, system outputs are normalized before comparison with the gold answer. For white-box evaluation, the top-\(K\) retrieved memory items returned by each system are used to compute retrieval and rank-based metrics. All LLM calls used in benchmark construction, including profile generation, dialogue generation, query generation, and memory-system backends that require an LLM, use \texttt{gpt-5.0-mini}.

\subsection{Main Results}\label{sec4.3}

We first report the main results for the six memory systems on MemConflict. The analysis is organized around two evaluation levels: black-box results measure answer correctness under dynamic, static, and conditional conflicts, while white-box results examine whether systems retrieve the gold memory item and how highly it is ranked. These results provide an overview of system effectiveness and memory-level behavior across the three conflict types.

\subsubsection{\textbf{Black-box Performance Across Conflict Types}}

Black-box evaluation measures the user-visible effectiveness of a memory system: after processing a user's multi-session interaction history, the system must answer conflict-targeted queries using its own memory mechanism. This perspective is important because the final response is the output exposed to users, regardless of how memory is internally stored, retrieved, or updated. We assess the returned answers using AA across dynamic, static, and conditional conflicts, and report UOCS and CRS as diagnostics for dynamic and static settings.

Table~\ref{tab2} reports the black-box evaluation results for the six memory systems. No system dominates all metrics and conflict types. MemOS achieves the highest average AA, with the best static and conditional AA and the highest UOCS, although LangMem obtains the highest dynamic AA. Letta and A-Mem show relatively strong average performance, but their strengths are concentrated in conditional conflicts rather than evenly distributed across conflict types. At the conflict-type level, static conflicts are the most difficult in terms of average AA, indicating that later contradictory mentions can substantially interfere with stable user information. Conditional conflicts obtain higher AA overall, but the results are highly polarized: MemOS, Letta, Mem0, and A-Mem perform well, whereas LangMem and Memobase perform much worse. The diagnostic metrics further show that final-answer correctness and conflict awareness are not equivalent. CRS remains low for all systems, with the best score only reaching 0.2501, and Memobase obtains relatively high static AA but the lowest CRS. This suggests that a system may return the correct stable value without explicitly recognizing the underlying contradiction.

These answer-level patterns suggest that different memory designs favor different forms of validity. Systems with stronger updating or consolidation mechanisms, such as LangMem, perform better on dynamic conflicts but do not necessarily preserve condition-value bindings, as reflected in their weaker conditional results. Systems with more explicit memory management, especially MemOS, appear more robust across static and conditional conflicts, suggesting that structured memory lifecycle control may help retain stable facts and preferences under competing memories. By contrast, profile-centric or lightweight memory designs show more uneven behavior: Memobase recovers some dynamic and static answers but has very low CRS and conditional AA, while Mem0 performs well on conditional conflicts but poorly on dynamic ones. Overall, temporal updating, factual preservation, and conditional applicability stress different aspects of long-term memory behavior, which explains why answer-level performance varies substantially across conflict types.

\begin{table}[t]
\centering
\caption{Black-box Performance of Memory Systems on MemConflict by Conflict Type.}
\renewcommand{\arraystretch}{1.15}
\label{tab2}
\begin{threeparttable}
\begin{tabular}{l@{\hspace{1.2em}}cc@{\hspace{1.2em}}cc@{\hspace{1.2em}}c@{\hspace{1.2em}}c}
\toprule
\multirow{2}{*}{\textbf{Memory System}}
& \multicolumn{2}{c}{\textbf{Dynamic}}
& \multicolumn{2}{c}{\textbf{Static}}
& \textbf{Conditional}
& \textbf{Average} \\
\cmidrule(lr){2-3}
\cmidrule(lr){4-5}
\cmidrule(lr){6-6}
\cmidrule(lr){7-7}
& \textbf{AA} & \textbf{UOCS}
& \textbf{AA} & \textbf{CRS}
& \textbf{AA}
& \textbf{AA} \\
\midrule
A-Mem    & 0.3596 & 0.2911 & 0.2639 & \textbf{0.2501} & 0.7122 & 0.4452 \\
LangMem  & \textbf{0.4966} & 0.3579 & 0.1944 & 0.2083 & 0.1556 & 0.2822 \\
Letta    & 0.3955 & 0.3527 & 0.2223 & 0.2031 & 0.8435 & 0.4871 \\
MemOS    & 0.3793 & \textbf{0.3818} & \textbf{0.4375} & 0.2361 & \textbf{0.8449} & \textbf{0.5539} \\
Mem0     & 0.1224 & 0.1130 & 0.1944 & 0.1528 & 0.7667 & 0.3612 \\
Memobase & 0.4058 & 0.3476 & 0.4167 & 0.0694 & 0.2434 & 0.3553 \\
\bottomrule
\end{tabular}
\begin{tablenotes}[flushleft]
\footnotesize
\item \textit{Note.} Bold values indicate the best result in each column.
\end{tablenotes}
\end{threeparttable}
\end{table}

\subsubsection{\textbf{White-box Memory Retrieval and Ranking}}

White-box evaluation examines whether a system retrieves the gold memory item needed to support the correct answer. For the same conflict-targeted queries used in the black-box evaluation, we analyze the top-3 memory items returned by each system and report SEH@3 and SRS. This perspective focuses on memory retrieval and ranking: a system may retrieve the gold memory item but rank it low in the retrieved list.

Table~\ref{tab3} reports the white-box results. MemOS achieves the highest average SEH@3 and SRS, indicating the strongest overall retrieval and ranking of gold memory items. Letta ranks second on average and performs especially well on conditional conflicts, where it obtains the highest SEH@3. A-Mem also shows strong conditional memory performance, while LangMem is clearly specialized toward dynamic conflicts, achieving the best dynamic SEH@3 and SRS but much weaker static and conditional scores. Memobase retrieves and ranks memory items competitively for dynamic and static conflicts, but its conditional performance is low. These patterns broadly align with the black-box results, suggesting that answer-level strengths often reflect whether the gold memory item is retrievable and highly ranked.

The gap between SEH@3 and SRS provides an additional diagnostic signal. Across most systems and conflict types, SEH@3 is higher than SRS, showing that the gold memory item may appear within the retrieved set but at a lower rank. This gap is especially important under conflict, because a low-ranked valid memory can be overshadowed by outdated, contradictory, conditionally inapplicable, or non-target memories during response generation. Static conflicts remain difficult at the memory level, with generally lower SEH@3 and SRS than the other conflict types, indicating that stable factual memories are hard to surface once later contradictory mentions are present. Conditional conflicts show the strongest system polarization: Letta, MemOS, A-Mem, and Mem0 retrieve condition-relevant memories effectively, whereas LangMem and Memobase fail to preserve or retrieve condition-value associations reliably. White-box analysis therefore complements black-box accuracy by showing whether systems make the gold memory item available and prominent enough for downstream answer generation.

\begin{table}[t]
\centering
\caption{White-box Memory Retrieval and Ranking of Memory Systems on MemConflict by Conflict Type.}
\renewcommand{\arraystretch}{1.15}
\label{tab3}
\begin{threeparttable}
\begin{tabular}{l@{\hspace{0.9em}}cc@{\hspace{0.9em}}cc@{\hspace{0.9em}}cc@{\hspace{0.9em}}cc}
\toprule
\multirow{2}{*}{\textbf{\begin{tabular}[c]{@{}c@{}}Memory\\System\end{tabular}}}
& \multicolumn{2}{c}{\textbf{Dynamic}}
& \multicolumn{2}{c}{\textbf{Static}}
& \multicolumn{2}{c}{\textbf{Conditional}}
& \multicolumn{2}{c}{\textbf{Average}} \\
\cmidrule(lr){2-3}
\cmidrule(lr){4-5}
\cmidrule(lr){6-7}
\cmidrule(lr){8-9}
& \textbf{SEH@3} & \textbf{SRS}
& \textbf{SEH@3} & \textbf{SRS}
& \textbf{SEH@3} & \textbf{SRS}
& \textbf{SEH@3} & \textbf{SRS} \\
\midrule
A-Mem    & 0.5205 & 0.4341 & 0.3611 & 0.2854 & 0.8111 & 0.7288 & 0.5642 & 0.4828 \\
LangMem  & \textbf{0.7842} & \textbf{0.7089} & 0.3194 & 0.2697 & 0.2012 & 0.1944 & 0.4349 & 0.3910 \\
Letta    & 0.5394 & 0.4620 & 0.4167 & 0.3099 & \textbf{0.9046} & 0.7653 & 0.6202 & 0.5124 \\
MemOS    & 0.5548 & 0.4552 & \textbf{0.5694} & \textbf{0.4886} & 0.8889 & \textbf{0.8198} & \textbf{0.6710} & \textbf{0.5879} \\
Mem0     & 0.2003 & 0.1587 & 0.2917 & 0.2401 & 0.8222 & 0.7780 & 0.4381 & 0.3923 \\
Memobase & 0.5925 & 0.5204 & 0.5278 & 0.4557 & 0.3021 & 0.2877 & 0.4741 & 0.4213 \\
\bottomrule
\end{tabular}
\begin{tablenotes}[flushleft]
\footnotesize
\item \textit{Note.} Bold values indicate the best result in each column.
\end{tablenotes}
\end{threeparttable}
\end{table}

\subsection{Ablation and Sensitivity Analysis}

We next analyze how controlled benchmark factors affect memory performance under conflict. Each analysis varies one factor while keeping the remaining settings fixed, covering retrieval depth, dialogue length, distractor injection, query formulation, and conflict distance. These sensitivity analyses test whether the observed weaknesses persist across evaluation conditions and identify which aspects of long-term memory make conflict-aware retrieval and response generation more difficult.

\subsubsection{\textbf{Effect of Retrieval Depth}}

We examine retrieval depth by varying the number of retrieved memory items used in white-box evaluation. The main setting uses \(K=3\); here we compare \(K=2\), \(K=3\), and \(K=5\) while keeping the query set, memory histories, and evaluation protocol unchanged. This analysis tests whether deeper retrieval improves the chance of retrieving the gold memory item and whether that item is ranked high enough to remain useful under conflict.

\begin{figure}[t]
\centering
\includegraphics[width=\textwidth]{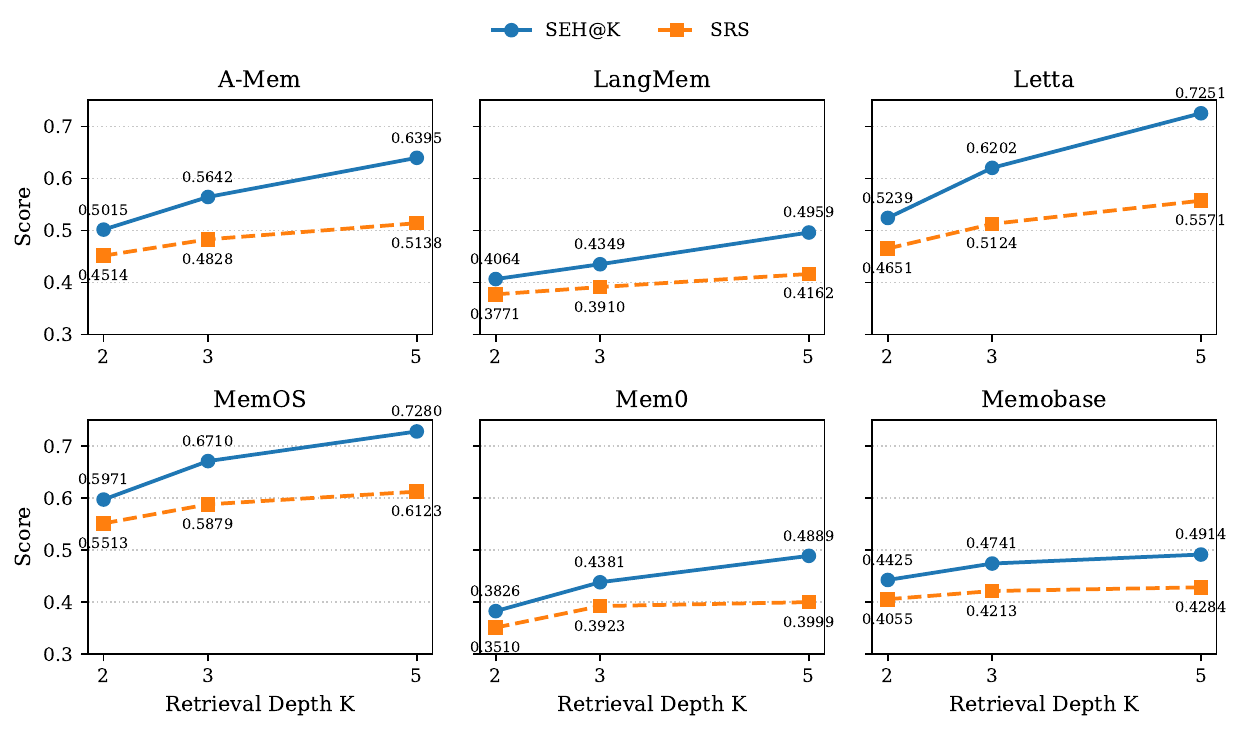}
\caption{Average SEH@K and SRS of memory systems under different retrieval depths.}
\label{fig4}
\end{figure}

Figure~\ref{fig4} shows the average SEH@K and SRS of each memory system under different retrieval depths. Increasing \(K\) improves both metrics for most systems, but the gains are uneven. Letta benefits most clearly, with SEH@K increasing from 0.5239 at \(K=2\) to 0.7251 at \(K=5\), while SRS rises from 0.4651 to 0.5571. MemOS also improves with deeper retrieval and remains the strongest system at \(K=5\), reaching 0.7280 in SEH@K and 0.6123 in SRS. A-Mem, LangMem, and Mem0 show more moderate improvements, whereas Memobase changes only slightly. These trends indicate that increasing retrieval depth can expose the gold memory item more often, but the benefit depends strongly on the memory system.

The different behavior of SEH@K and SRS suggests that retrieval depth mainly improves memory-item coverage rather than fully resolving ranking quality. Additional retrieved items increase the chance that the gold memory item appears in the retrieved set, but they do not necessarily place it near the top. Memobase is particularly informative: its conditional-conflict SEH@K and SRS remain unchanged across \(K=2\), \(K=3\), and \(K=5\), indicating that the bottleneck is not shallow retrieval depth alone. Condition-relevant memories may instead be absent from the retrievable candidate set or insufficiently preserved in the memory representation. Robust conflict-aware memory therefore requires not only retrieving more items, but also preserving valid memories and ranking them above outdated, contradictory, or conditionally inapplicable alternatives.

\subsubsection{\textbf{Effect of Dialogue Length}}

We examine the effect of dialogue length by constructing a longer-history version of the evaluation data. Starting from the default dialogue histories, we expand each user history with additional unrelated but conversationally coherent QA turns adapted from the custom-history construction setting in LongMemEval\footnote{\url{https://github.com/xiaowu0162/LongMemEval}.} and supplemented with LLM-generated unrelated QA turns covering general explanatory and math-related topics. These supplementary turns are inserted either into existing sessions or as additional sessions, while the original conflict instances, gold answers, and query protocol are kept unchanged. This design increases the amount of intervening context around memory-relevant information without changing the target conflict structure, allowing us to isolate the effect of longer dialogue histories.

\begin{figure}[t]
\centering
\includegraphics[width=\textwidth]{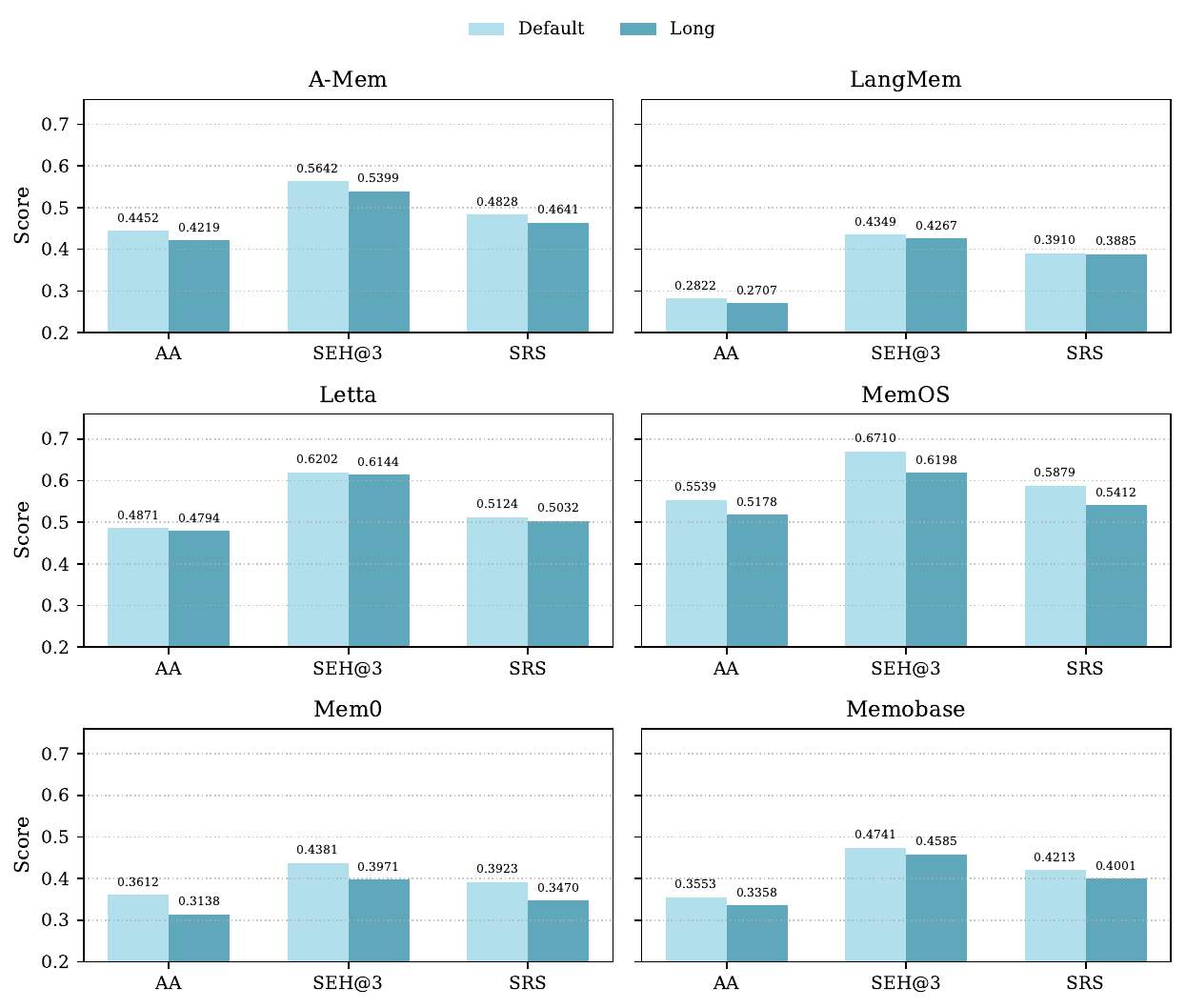}
\caption{Average AA, SEH@3, and SRS of memory systems under different dialogue lengths.}
\label{fig5}
\end{figure}

Figure~\ref{fig5} compares the average AA, SEH@3, and SRS of each memory system under the default and longer dialogue settings. Longer histories reduce performance for all evaluated systems, but the magnitude of degradation varies. MemOS remains the strongest system under the longer setting, although all three metrics decrease. Letta is comparatively stable, whereas Mem0 shows the largest drop in average AA and SRS. A-Mem and Memobase exhibit moderate declines, while LangMem changes only slightly in white-box metrics but remains low in average AA. These results indicate that longer dialogue histories make conflict-aware memory use more difficult, with different systems showing different levels of robustness to additional context.

The degradation under longer histories suggests that conflict difficulty depends not only on the presence of competing memories, but also on how much surrounding context accumulates around relevant memory items. Additional unrelated turns can dilute target memories, increase retrieval competition, and make valid memories harder to surface and rank highly. The simultaneous decline in black-box and white-box metrics indicates that longer histories affect both final-answer generation and memory retrieval quality. Systems with stronger memory organization or consolidation appear less sensitive to irrelevant intervening context, suggesting that long-term memory evaluation should vary dialogue length rather than rely on a single fixed-history setting.

\begin{figure}[t]
\centering
\includegraphics[width=\textwidth]{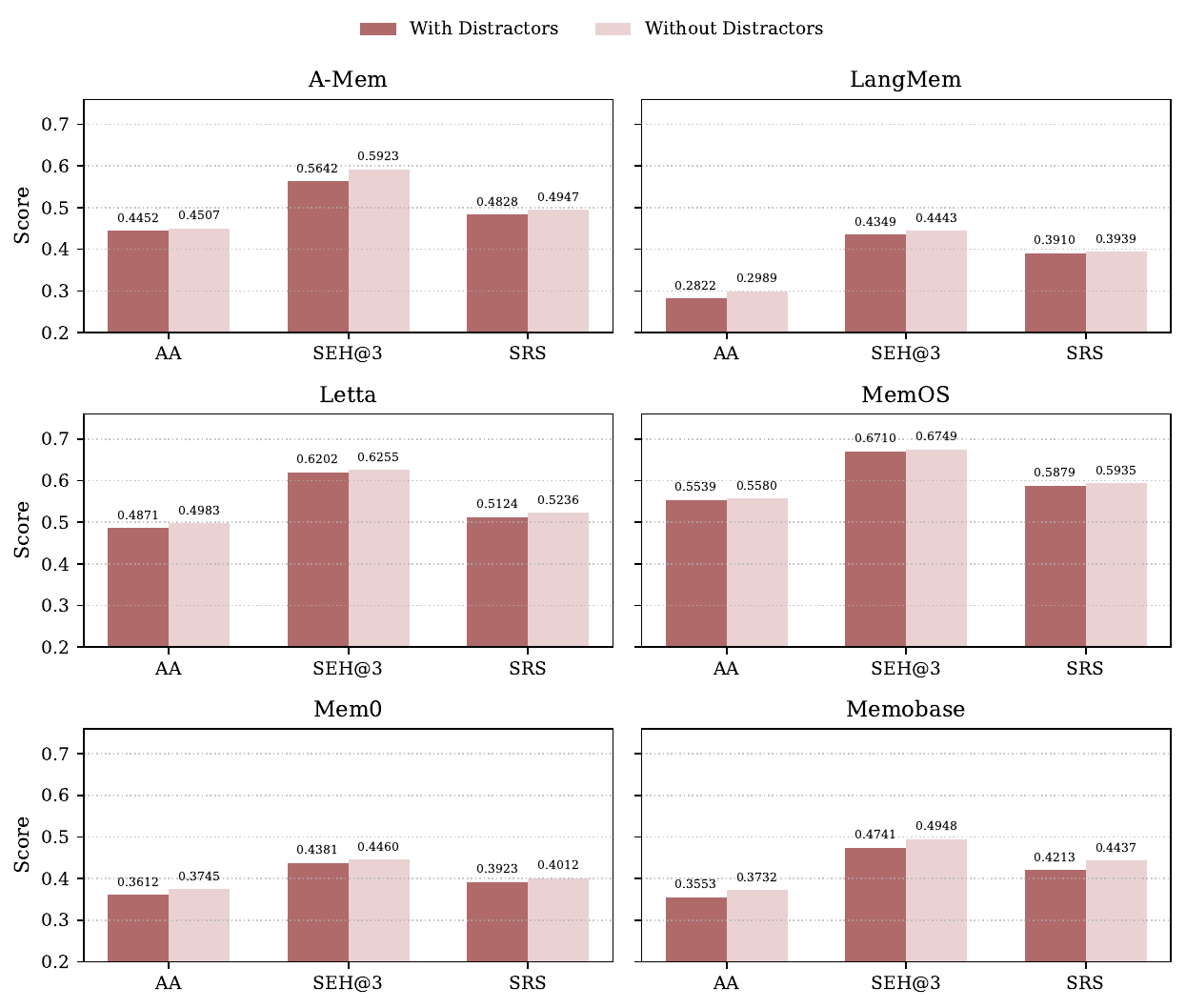}
\caption{Average AA, SEH@3, and SRS of memory systems with and without distractor injection.}
\label{fig6}
\end{figure}

\subsubsection{\textbf{Effect of Distractor Injection}}

We next examine the effect of semantically similar distractors. In the default MemConflict setting, interaction histories include non-target information from related entities, creating competition with the target memory during retrieval and response generation. To isolate this factor, we compare the default setting with a variant that removes distractor information while keeping the original conflict instances, gold answers, query protocol, and evaluation metrics unchanged. This comparison measures how much semantically similar but non-target memory content affects conflict-aware memory use.

Figure~\ref{fig6} compares the average AA, SEH@3, and SRS of each memory system with and without distractor injection. Removing distractors improves performance for all systems, although the gains vary. MemOS remains the strongest system in both settings and shows only a modest improvement after distractor removal, suggesting relatively strong robustness to semantically similar interference. Memobase and A-Mem show more visible gains in white-box metrics, indicating that distractors affect their ability to retrieve and rank the gold memory item. LangMem and Mem0 also improve, but their absolute performance remains limited compared with stronger systems. These results show that distractors introduce measurable difficulty, but they do not fully explain the performance gaps among memory systems.

The effect of distractors suggests that conflict-aware memory use depends on more than topical relevance. When related entities or similar attributes appear in the history, systems may retrieve plausible but non-target memories or rank them above the gold memory item. The improvement in both AA and white-box metrics after distractor removal indicates that distractors affect final responses as well as memory retrieval and ranking. However, performance differences persist even without distractors, suggesting that robust memory conflict handling also requires systems to preserve target-entity boundaries, maintain attribute-specific memories, and select memory that is valid for the current query.

\subsubsection{\textbf{Effect of Query Formulation}}

We examine the effect of query formulation by evaluating systems under implicit queries. Unlike direct queries, which explicitly mention the target memory attribute, implicit queries require the system to infer the relevant memory from less explicit wording. The histories, conflict instances, gold answers, and evaluation protocol are kept unchanged, so the comparison isolates the effect of query wording on memory access and use. This setting tests whether systems can retrieve and apply valid memory when the query does not directly match the surface form of stored information.

Table~\ref{tab4} reports the results under implicit query formulation. Compared with the direct-query setting, implicit queries generally reduce both black-box and white-box performance, although the effect varies across systems and conflict types. MemOS achieves the best average AA, SEH@3, and SRS, indicating relatively strong robustness to less explicit query wording. Letta and A-Mem also maintain strong performance, especially on conditional conflicts. LangMem remains strongest on dynamic conflicts, but its static and conditional results remain low, suggesting that implicit wording does not change its sensitivity to non-temporal conflict types. Memobase performs competitively on static conflicts but remains weak on conditional conflicts, while Mem0 retains strong conditional performance but degrades substantially on dynamic conflicts.

These results suggest that query formulation affects how systems access and use memory. With implicit queries, systems cannot rely only on lexical overlap between the query and stored memory items; they must infer the target attribute, retrieve semantically appropriate memories, and map those memories to the intended answer. This is especially difficult under conflict, where semantically related but invalid memories may appear plausible. The decreases in SEH@3 and SRS indicate that implicit wording makes the gold memory item harder to retrieve and rank highly, while the decrease in AA shows that this retrieval difficulty also affects final answers. Evaluating only direct queries may therefore overestimate memory robustness in realistic user interactions.

\begin{table}[t]
\centering
\caption{Performance of Memory Systems under Implicit Query Formulation by Conflict Type.}
\renewcommand{\arraystretch}{1.35}
\label{tab4}
\begin{threeparttable}
\resizebox{\textwidth}{!}{%
\begin{tabular}{lcccccccccccc}
\toprule
\multirow{2}{*}{\textbf{\begin{tabular}[c]{@{}c@{}}Memory\\System\end{tabular}}}
& \multicolumn{3}{c}{\textbf{Dynamic}}
& \multicolumn{3}{c}{\textbf{Static}}
& \multicolumn{3}{c}{\textbf{Conditional}}
& \multicolumn{3}{c}{\textbf{Average}} \\
\cmidrule(lr){2-4}
\cmidrule(lr){5-7}
\cmidrule(lr){8-10}
\cmidrule(lr){11-13}
& \textbf{AA} & \textbf{SEH@3} & \textbf{SRS}
& \textbf{AA} & \textbf{SEH@3} & \textbf{SRS}
& \textbf{AA} & \textbf{SEH@3} & \textbf{SRS}
& \textbf{AA} & \textbf{SEH@3} & \textbf{SRS} \\
\midrule
A-Mem    & 0.3356 & 0.4812 & 0.3899 & 0.2231 & 0.3550 & 0.2799 & 0.7026 & 0.8023 & 0.7189 & 0.4204 & 0.5462 & 0.4629 \\
LangMem  & \textbf{0.4795} & \textbf{0.7671} & \textbf{0.7013} & 0.1493 & 0.2750 & 0.2406 & 0.1475 & 0.1956 & 0.1833 & 0.2588 & 0.4126 & 0.3751 \\
Letta    & 0.3943 & 0.5291 & 0.4445 & 0.2127 & 0.3472 & 0.3028 & 0.8221 & \textbf{0.8836} & 0.7605 & 0.4764 & 0.5866 & 0.5026 \\
MemOS    & 0.3399 & 0.4709 & 0.3792 & \textbf{0.4306} & 0.5139 & \textbf{0.4828} & \textbf{0.8267} & 0.8667 & \textbf{0.8102} & \textbf{0.5324} & \textbf{0.6172} & \textbf{0.5574} \\
Mem0     & 0.0916 & 0.1884 & 0.1425 & 0.1875 & 0.2361 & 0.1882 & 0.7525 & 0.8004 & 0.7479 & 0.3439 & 0.4083 & 0.3595 \\
Memobase & 0.3866 & 0.5890 & 0.5124 & 0.4061 & \textbf{0.5194} & 0.4447 & 0.2361 & 0.2967 & 0.2834 & 0.3429 & 0.4684 & 0.4135 \\
\bottomrule
\end{tabular}%
}
\begin{tablenotes}[flushleft]
\footnotesize
\item \textit{Note.} Values are reported for implicit queries; bold values indicate the best result in each column.
\end{tablenotes}
\end{threeparttable}
\end{table}

\subsubsection{\textbf{Effect of Conflict Distance}}

We examine the effect of conflict distance by varying how far apart competing memory candidates are distributed in the interaction history. In the near-distance setting, competing candidates are separated by 5--10 intervening sessions; in the far-distance setting, they are separated by 20--25 intervening sessions. The default setting corresponds to the main evaluation setting, where conflict distances are sampled from a broader range rather than fixed to a specific near or far interval. The conflict instances, query protocol, and evaluation metrics are kept unchanged, allowing us to isolate how temporal separation between competing candidates affects memory use.

\begin{figure}[t]
\centering
\includegraphics[width=\textwidth]{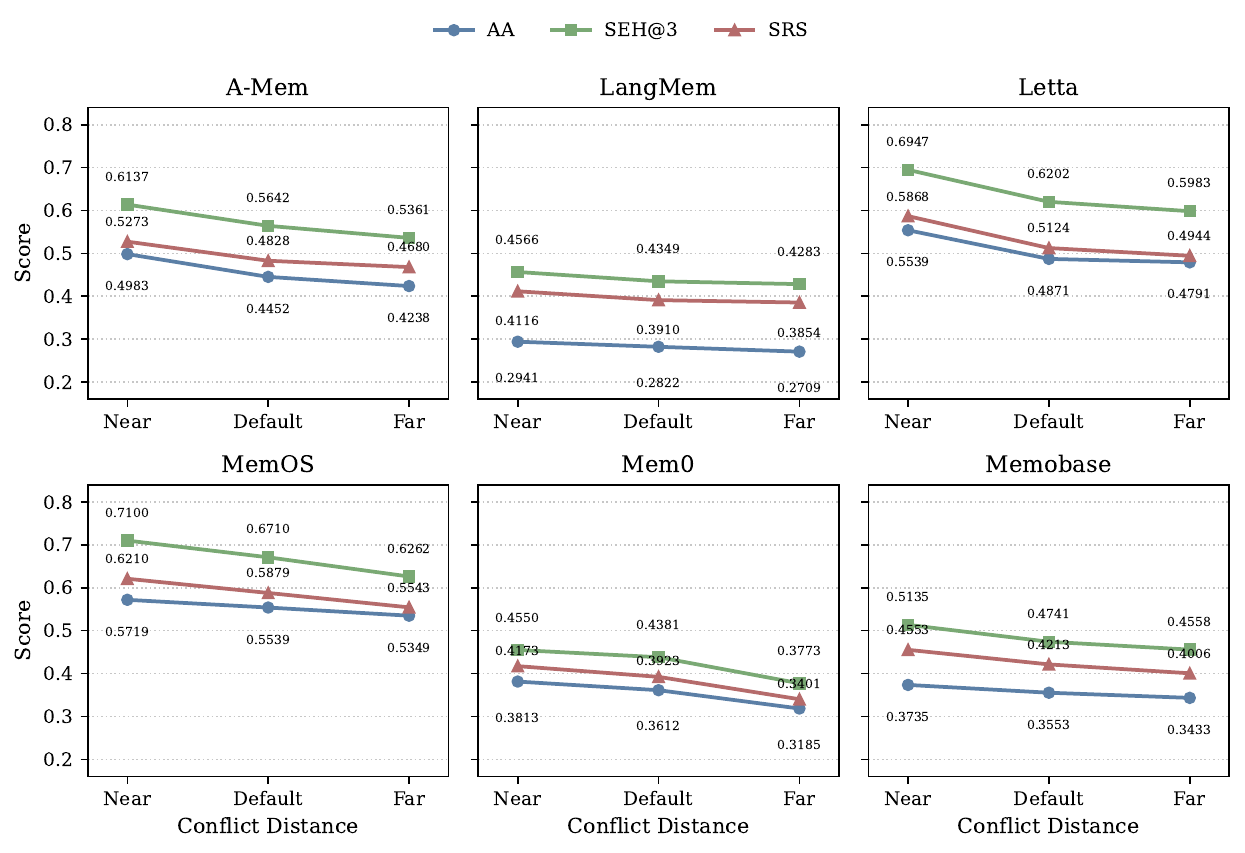}
\caption{Average AA, SEH@3, and SRS of memory systems under different conflict distances.}
\label{fig7}
\end{figure}

Figure~\ref{fig7} shows the average AA, SEH@3, and SRS of each memory system under the near, default, and far conflict-distance settings. Performance decreases as conflict distance increases across all systems. This trend appears in both black-box and white-box metrics, indicating that larger separation between competing candidates makes it harder to produce the correct answer and to retrieve and rank the gold memory item. MemOS and Letta maintain relatively strong absolute performance across distances, but both decline from near to far settings. Mem0 is especially sensitive to increased distance, with noticeable drops in all three metrics. LangMem changes more gradually, but its absolute AA remains low. These results suggest that conflict distance is an important source of difficulty for long-term memory systems.

The degradation suggests that conflict resolution depends on connecting distributed memories across the interaction history. When competing memories are separated by only 5--10 sessions, systems may still retrieve them within a similar local memory neighborhood or preserve their relationship during memory updating. When the distance grows to 20--25 intervening sessions, relevant memories become more dispersed and separated by unrelated sessions, making it harder to determine which memory remains valid for the current query. The simultaneous decline in SEH@3 and SRS shows that distance affects both retrieval coverage and ranking quality, while the decline in AA indicates that these retrieval-level difficulties also affect final answers. Robust memory systems should therefore avoid relying only on local recency or short-range retrieval, and should maintain links between related memory candidates across longer temporal gaps.

\subsection{System-Level Diagnostic Analysis}

Beyond effectiveness under controlled benchmark factors, we further examine system-level properties that affect the practical use of long-term memory systems. Using the same dataset and default configuration as the main evaluation in Section~\ref{sec4.3}, we focus on two aspects: runtime efficiency across memory addition, retrieval, and response generation, and reliability diagnosis that separates retrieval failures from failures to use retrieved valid memories. This analysis complements the preceding performance results by showing not only how well systems answer conflict-targeted queries, but also where computational cost and failure sources arise.

\subsubsection{\textbf{Efficiency Analysis}}

We analyze runtime efficiency to understand the practical cost of different memory designs beyond accuracy and memory-level performance. Table~\ref{tab5} reports memory addition time, retrieval time, response time, and total runtime for each evaluated system. The results show substantial differences in both total cost and cost distribution. MemOS achieves the lowest total runtime, with the lowest addition time and response time. A-Mem has the lowest retrieval time, but its total runtime remains high because of expensive memory addition. Mem0 has by far the highest total runtime, dominated by memory addition, while Letta and Memobase incur relatively higher retrieval or response costs. These results show that memory systems differ not only in conflict-handling effectiveness, but also in where computational cost is concentrated.

These efficiency differences should be interpreted in light of each system's memory architecture and deployment assumptions. Low retrieval latency does not necessarily imply stronger retrieval quality: for example, A-Mem retrieves quickly largely because its memories are stored and searched locally, while its addition stage still incurs substantial cost from organizing and linking memory units. Systems that perform more memory construction, extraction, consolidation, or external service calls during ingestion can shift cost toward the addition stage, whereas systems that defer more processing until query time may incur higher retrieval or response latency. The results therefore highlight a practical trade-off in long-term memory design: memory systems must balance the cost of writing and organizing memory against the quality and latency of later retrieval and response generation.

\begin{table}[t]
\centering
\caption{Runtime Efficiency of Memory Systems.}
\renewcommand{\arraystretch}{1.15}
\label{tab5}
\begin{threeparttable}
\begin{tabular}{lrrrr}
\toprule
\textbf{Memory System} & \textbf{Addition Time} & \textbf{Retrieval Time} & \textbf{Response Time} & \textbf{Total Runtime} \\
\midrule
A-Mem    & 9213.7727  & \textbf{2.2264}   & 906.9217  & 10122.9208 \\
LangMem  & 638.8427   & 159.5139  & 878.0856  & 1676.4422 \\
Letta    & 2572.2743  & 368.4233  & 992.3163  & 3933.0139 \\
MemOS    & \textbf{472.6703} & 69.7730   & \textbf{813.8600} & \textbf{1356.3033} \\
Mem0     & 40215.7970 & 89.0861   & 905.9518  & 41210.8349 \\
Memobase & 1266.6696  & 471.1545  & 1344.0923 & 3081.9164 \\
\bottomrule
\end{tabular}
\begin{tablenotes}[flushleft]
\footnotesize
\item \textit{Note.} All times are measured in seconds; total runtime is the sum of addition, retrieval, and response time. Lower values indicate better efficiency. Bold values indicate the lowest runtime in each column.
\end{tablenotes}
\end{threeparttable}
\end{table}

\subsubsection{\textbf{Reliability and Failure Diagnosis}}

We further diagnose reliability by connecting answer correctness with memory-level retrieval outcomes. The preceding analyses report black-box performance and white-box memory retrieval and ranking quality separately; here, we examine whether a retrieved gold memory item is actually converted into a correct final answer. We also decompose incorrect cases into retrieval failures, where the gold memory item is absent from the retrieved set, and utilization failures, where the gold memory item is retrieved but the final answer is still incorrect. This diagnosis separates retrieval-side limitations from memory-utilization limitations that are otherwise conflated in answer-only evaluation.

We define the Evidence Utilization Gap (EUG) as the difference between SEH@3 and AA, which approximates the aggregate gap between retrieving the gold memory item and converting it into a correct final answer. Table~\ref{tab6} reports EUG across conflict types. A larger EUG indicates a stronger mismatch between memory retrieval and answer generation, rather than lower performance by itself. LangMem shows the largest average EUG, mainly due to its high dynamic gap, suggesting that it can surface relevant updated memories but does not always convert them into temporally valid answers. Letta exhibits the largest static EUG, indicating that its persistent memory state may expose useful memories while still failing to resolve contradictory mentions correctly. Mem0 has the smallest average EUG, but this should be interpreted together with its lower retrieval and answer performance: a small gap may reflect fewer cases in which the gold memory item is retrieved, rather than uniformly stronger reliability.

\begin{table}[t]
  \centering
  \caption{Evidence Utilization Gaps by Conflict Type.}
  \renewcommand{\arraystretch}{1.15}
  \label{tab6}
  \begin{threeparttable}
  \begin{tabular}{lcccc}
    \toprule
    \textbf{Memory System} & \textbf{Dynamic EUG} & \textbf{Static EUG} & \textbf{Conditional EUG} & \textbf{Average EUG} \\
    \midrule
    A-Mem    & 0.1609 & 0.0972 & \textbf{0.0989} & 0.1190 \\
    LangMem  & \textbf{0.2876} & 0.1250 & 0.0456 & \textbf{0.1527} \\
    Letta    & 0.1439 & \textbf{0.1944} & 0.0611 & 0.1331 \\
    MemOS    & 0.1755 & 0.1319 & 0.0440 & 0.1171 \\
    Mem0     & 0.0779 & 0.0973 & 0.0555 & 0.0769 \\
    Memobase & 0.1867 & 0.1111 & 0.0587 & 0.1188 \\
    \bottomrule
  \end{tabular}
\begin{tablenotes}[flushleft]
  \footnotesize
\item \textit{Note.} EUG denotes the Evidence Utilization Gap, defined as the proportion of queries for which the gold memory item is retrieved but the final answer remains incorrect. Larger values indicate stronger memory--answer mismatch; bold values indicate the largest gap in each column.
\end{tablenotes}
  \end{threeparttable}
\end{table}

\begin{figure}[t]
  \centering
  \includegraphics[width=\textwidth]{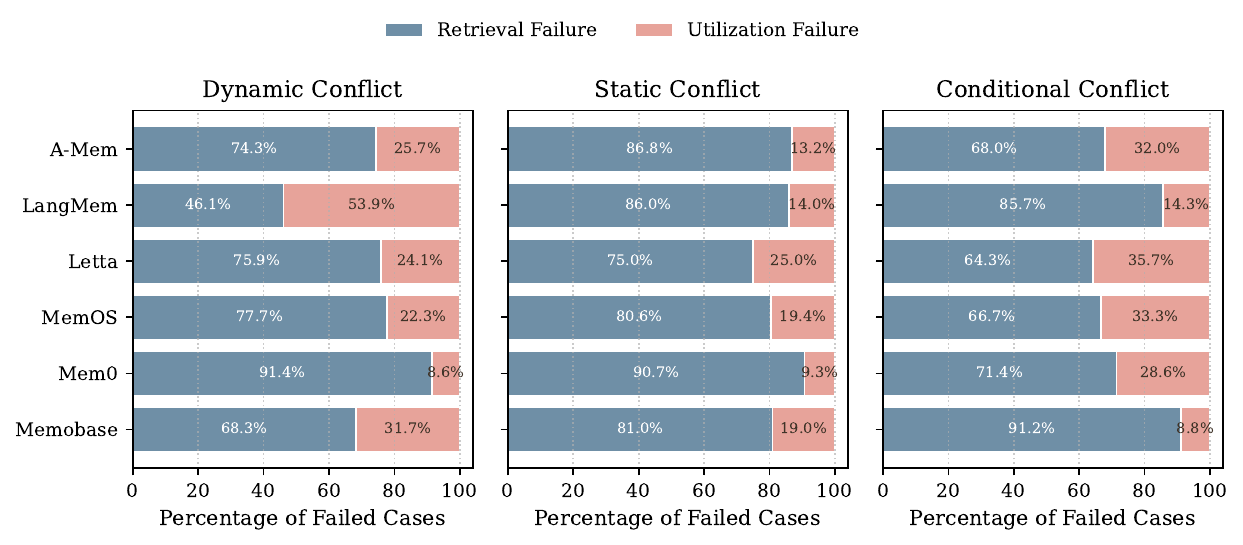}
  \caption{Retrieval and utilization failures across conflict types.}
  \label{fig8}
\end{figure}

Figure~\ref{fig8} further decomposes incorrect cases into retrieval failures and utilization failures. Across most systems and conflict types, retrieval failures account for the dominant share of errors, showing that many incorrect answers occur when the gold memory item is not present in the retrieved set. The figure also reveals non-trivial utilization problems. LangMem under dynamic conflicts has a larger share of utilization failures than retrieval failures, suggesting that temporal updating remains difficult even when the relevant memory is available. Conditional conflicts also show relatively larger utilization-failure shares for A-Mem, Letta, MemOS, and Mem0, indicating that context-dependent memory requires not only retrieval but also correct use of value-condition associations during response generation. These findings show that MemConflict exposes different reliability bottlenecks across systems, rather than treating all incorrect answers as the same type of failure.

\subsection{Suggestions for Optimizing Memory Systems}

The experimental results provide several implications for the design of long-term memory systems under conflicting information. MemOS shows the strongest overall performance and efficiency, while Letta is competitive on conditional conflicts and LangMem is strongest on dynamic conflicts. However, systems that perform well in one setting often degrade in others: LangMem remains weak on static and conditional conflicts, Memobase struggles with conditional conflicts and contradiction awareness, and Mem0 is sensitive to temporal changes and longer histories. Across black-box and white-box evaluation, no system dominates all metrics and conflict types, indicating that conflict-aware memory requires more than generic memory retrieval. Dynamic conflicts expose weaknesses in temporal update modeling, static conflicts reveal difficulty in handling contradictory mentions that should not overwrite stable facts, and conditional conflicts show that systems must preserve the applicability conditions of remembered preferences or values. The sensitivity analyses further show that performance degrades when relevant memories are buried in longer histories, mixed with semantically similar distractors, expressed through implicit queries, or separated by larger conflict distances. The diagnostic analysis also shows that failures arise from different sources: many errors are caused by missing gold memory items during retrieval, while some systems retrieve the relevant memory but fail to use it correctly during generation.

These findings suggest several practical directions for improving memory systems. First, memory representations should explicitly encode temporal state, source attribution, and applicability conditions, rather than storing extracted facts as isolated snippets. Such metadata can help systems decide whether a memory is currently valid, contradicted, attributable to the target user, or conditionally applicable. Second, retrieval should be complemented with conflict-aware reranking mechanisms that rank memory items according to temporal validity, factual consistency, condition matching, and query relevance, especially when distractors or distant conflicts are present. Third, memory systems should introduce memory-verification steps before response generation, so that retrieved valid memories are not ignored or overridden by more salient but invalid memories. Finally, optimization should consider efficiency together with reliability: more elaborate memory construction may improve organization in some systems, but it can also introduce substantial writing cost, so practical systems need to balance memory writing, retrieval quality, and answer-time reasoning. Overall, the results suggest that robust long-term memory systems should be designed as evidence-aware and conflict-aware pipelines, rather than as passive stores of accumulated user information.

\section{Conclusion and Future Work}\label{sec5}

This paper proposes \textbf{MemConflict}, a diagnostic evaluation framework for long-term memory systems operating under conflicting information. MemConflict evaluates whether systems can identify temporally valid updates, preserve factually correct information despite contradictory mentions, and select contextually applicable memory under long-horizon, multi-session interactions. To support this goal, we constructed a controlled benchmark that instantiates dynamic, static, and conditional conflicts together with semantically similar distractors, and evaluated memory systems from both black-box and white-box perspectives. The results show that current systems exhibit uneven strengths across conflict types, and that answer correctness does not always align with memory retrieval and ranking quality. By connecting final answers with memory-level behavior, MemConflict provides a more fine-grained view of long-term memory reliability than answer-only evaluation.

Despite these contributions, MemConflict has several limitations. Although the benchmark is designed to be realistic and internally coherent, it is still constructed through controlled simulation rather than collected from naturally occurring long-term human interactions. As a result, it captures important conflict structures systematically, but cannot fully represent the variability and ambiguity of real-world conversational memory. In addition, the current benchmark focuses on a specific set of memory difficulties, namely temporal updates, factual contradictions, and condition-dependent preferences, while other challenges such as multimodal memory grounding, richer social interaction, or strategically omitted information are not explicitly modeled. Finally, our experiments evaluate a representative but limited set of memory systems, and the current white-box protocol assumes that memory items supporting the gold answer can be explicitly retrieved and ranked, which may be difficult for opaque or proprietary memory systems in deployed settings.

These limitations suggest several directions for future work. A natural extension is to broaden MemConflict to more realistic and diverse interaction settings, including multilingual conversations, multimodal memory, and memory involving multiple interacting users or agents. Another direction is to incorporate naturally occurring long-term interaction data while preserving the interpretability and diagnostic control of the benchmark. Future work may also expand the conflict taxonomy and query space by introducing richer forms of temporal change, factual contradiction, condition-dependent memory, nested or overlapping conflicts, and tasks that require systems to justify why competing memories are invalid in the current context. More broadly, MemConflict points toward a shift in long-term memory evaluation from answer-only benchmarking to memory-aware and conflict-aware analysis.

\begin{acks}
This work was supported by the National Natural Science Foundation of China (Grant No. 72271233), the National Social Science Fund of China (Grant No. 25CTQ022), the Fundamental Research Funds for the Central Universities (2024110591), Suzhou Key Laboratory of Artificial Intelligence and Social Governance Technologies (SZS2023007), Smart Social Governance Technology and Innovative Application Platform (YZCXPT2023101), and the Innovation System of the Integration between Industry and Education for Smart Governance (CJRH2024101).
\end{acks}

\bibliographystyle{ACM-Reference-Format}
\bibliography{sample-base}

\clearpage

\appendix
\setcounter{figure}{0}
\renewcommand{\thefigure}{A\arabic{figure}}

\section{Prompt Templates for MemConflict Construction and Evaluation}\label{sec_a1}

This appendix presents refined prompt templates used in MemConflict construction and evaluation. They are condensed from the implementation prompts for readability while preserving the main inputs, constraints, and output requirements. The complete prompts and implementation scripts are provided in the project repository.

\begin{figure}[h]
\centering
\fbox{
\begin{minipage}{0.96\linewidth}
\small
\textbf{Prompt Template: Structured User Profile Initialization}

You are a persona modeling expert constructing realistic long-term user profiles for a memory benchmark.

\textbf{Input:} a persona seed $z_u$ and predefined profile schemas.

\textbf{Task:} generate or refine a structured profile
$U_u=(I_u,D_u,C_u,T_u,R_u)$, including invariant attributes, dynamic status attributes, conditional preferences, personality traits and life goals, and related-entity information.

\textbf{Constraints:}
(1) preserve the predefined JSON schema;
(2) ensure all fields are realistic, internally consistent, and compatible with the persona seed;
(3) keep invariant facts stable, dynamic attributes plausible, and conditional preferences expressed as condition-value associations;
(4) generate related entities with realistic backgrounds and preferences that may later serve as semantically similar distractors;
(5) do not introduce unrelated or implausible information.

\textbf{Output:} valid JSON containing the completed structured user profile.
\end{minipage}
}
\caption{Prompt template for structured user profile initialization.}
\label{fig:prompt_profile}
\end{figure}

\begin{figure}[h]
\centering
\fbox{
\begin{minipage}{0.96\linewidth}
\small
\textbf{Prompt Template: Conflict and Distractor Construction}

You are constructing memory-conflict instances for a long-term memory benchmark.

\textbf{Input:} structured user profile, candidate user facts, conditional preference groups, related-entity facts, and available session positions.

\textbf{Task:} construct conflict instances and distractors for target attributes.

\textbf{Rules:}
(1) dynamic conflicts must represent true user-state updates, where the later value is valid after the update session;
(2) static conflicts must use an invariant user fact as the true value and a plausible but false contradictory mention as the competing value;
(3) conditional conflicts must preserve condition-value bindings, where multiple values are valid under different conditions but only one applies to a query;
(4) distractors must come from related entities, remain attributable to those entities, and be semantically similar to the target memory without changing the target user's gold value;
(5) all conflict values must be inserted into valid session positions without using future information.

\textbf{Output:} valid JSON specifying conflict type, target attribute, inserted memory items, distractors, session positions, and gold value or gold condition.
\end{minipage}
}
\caption{Prompt template for conflict and distractor construction.}
\label{fig:prompt_conflict}
\end{figure}

\begin{figure}[t]
\centering
\fbox{
\begin{minipage}{0.96\linewidth}
\small
\textbf{Prompt Template: Multi-Session Dialogue Generation}

You are writing one natural multi-turn dialogue session for a long-term memory benchmark.

\textbf{Input:} session type $\tau_t$, session outline, event types, dynamic updates $\Delta_t$, static information $\mathcal{P}_t$, conditional bindings $\mathcal{B}_t$, distractors $\mathcal{E}_t$, and the required number of turns.

\textbf{Task:} first form a concise session synopsis, then realize it as a coherent user-assistant dialogue.

\textbf{Constraints:}
(1) every non-empty item in $\Delta_t$, $\mathcal{P}_t$, $\mathcal{B}_t$, and $\mathcal{E}_t$ must be explicitly expressed;
(2) target-user memories and related-entity distractors must be correctly attributed;
(3) information should be distributed naturally across turns rather than listed mechanically;
(4) do not introduce unsupported facts, future-session information, or unrelated major life changes;
(5) the conversation should be realistic, coherent, and suitable for later memory retrieval.

\textbf{Output:} valid JSON containing the multi-turn dialogue.
\end{minipage}
}
\caption{Prompt template for session synopsis and dialogue generation.}
\label{fig:prompt_dialogue}
\end{figure}

\begin{figure}[t]
\centering
\fbox{
\begin{minipage}{0.96\linewidth}
\small
\textbf{Prompt Template: Query and Label Refinement}

You are refining benchmark questions and answers for naturalness while preserving their exact meaning.

\textbf{Input:} history prefix $H_t$, template questions, gold answers, conflict type, ability target, and difficulty label.

\textbf{Task:} rewrite each question and answer so that it sounds natural, clear, and realistic while preserving the original label.

\textbf{Constraints:}
(1) do not add facts or use future-session information;
(2) do not change the gold answer, conflict type, ability target, or difficulty;
(3) remove benchmark-internal wording such as point labels, raw field names, session indices, or template language;
(4) for ordinary static-conflict queries, keep the question neutral and let the answer state the inconsistency and the correct value when needed;
(5) for diagnostic conflict-awareness queries, ask directly whether incompatible candidates coexist in the available history;
(6) for dynamic conflicts, preserve the query-time interpretation of the latest valid state;
(7) for conditional conflicts, express condition-value associations as natural preference questions and answers.

\textbf{Output:} valid JSON containing the refined question, answer, conflict type, ability target, and difficulty.
\end{minipage}
}
\caption{Prompt template for query and label refinement.}
\label{fig:prompt_query}
\end{figure}

\begin{figure}[t]
\centering
\fbox{
\begin{minipage}{0.96\linewidth}
\small
\textbf{Prompt Template: LLM-Assisted Answer and Memory Matching}

You are a strict evaluator for memory-conflict experiments. Use the reference answer as the gold standard. Use the retrieved memory items only for memory-level matching.

\textbf{Input:} conflict type, question, reference answer $y_i^*$, model answer $\hat{y}_i$, and top-$K$ retrieved memory items $R_i^K$.

\textbf{Task:} judge whether the model answer is correct and identify the rank of the first retrieved memory item that supports the gold answer.

\textbf{Scoring Rules:}
(1) set answer correctness to 1 if the model answer is semantically consistent with the reference answer, and 0 otherwise;
(2) for dynamic conflicts, also judge whether the answer recognizes a true update and preserves the correct old-to-new temporal order;
(3) for static conflicts, also judge whether the answer recognizes incompatible candidates when the query asks about contradiction awareness;
(4) for conditional conflicts, judge whether the answer recovers the correct condition associated with the queried preference value;
(5) set the support rank to the 1-based rank of the first retrieved memory item whose content supports the gold answer, or 0 if no such item appears in top-$K$.

\textbf{Output:} valid JSON containing answer correctness, optional diagnostic indicators, support rank, and a short explanation.
\end{minipage}
}
\caption{Prompt template for LLM-assisted answer and memory-item matching.}
\label{fig:prompt_eval}
\end{figure}

\end{document}